\documentclass[amsmath,amssymb,aps,prd,11pt,tightenlines,superscriptaddress,nofootinbib,preprintnumbers,notitlepage]{revtex4-2}

\usepackage{xspace}
\usepackage{siunitx}
\usepackage{array}
\usepackage{tabularx}
\usepackage{xcolor}
\usepackage[mathlines]{lineno}
\usepackage{amsmath,amssymb,amsthm,amsfonts}
\usepackage{graphicx,tabularx}
\usepackage{color}
\usepackage{multirow}
\usepackage{comment}
\usepackage{enumitem}
\usepackage{subfigure}
\usepackage{orcidlink}
\usepackage{cleveref}
\usepackage{slashed}
\usepackage{scalerel}
\usepackage{wrapfig}
\usepackage{cancel}
\usepackage{xcolor}
\usepackage{multirow}
\usepackage{setspace}

\newcommand{\be}{\begin{equation} \begin{aligned}}
\newcommand{\ee}{\end{aligned} \end{equation}}
\newcommand{\beqa}{\begin{eqnarray}}
\newcommand{\eeqa}{\end{eqnarray}}

\def\figureautorefname~#1\null{Fig.\,#1\null}
\def\tableautorefname~#1\null{Tab.\,#1\null}
\def\equationautorefname~#1\null{Eq.\,(#1)\null}

\crefname{section}{Sec.}{Secs.}
\crefname{figure}{Fig.}{Figs.}
\crefname{equation}{Eq.}{Eqs.}
\crefname{table}{Table}{Tables}
\crefname{appendix}{Appendix}{Appendices}

\usepackage{subcaption}

\makeatletter
\renewcommand{\p@subsection}{}
\renewcommand{\p@subsubsection}{}
\makeatother
\begin{document}
%\linenumbers
% \preprint{CERN-EP-202X-0XX}
% \preprint{XX Month 202X}

\title{Optimization of muon suppression using sweeper magnets for the Forward Physics Facility at the HL-LHC
}

\author{Akitaka Ariga\,\orcidlink{0000-0002-6832-2466}}
\thanks{Email: akitaka.ariga@cern.ch}
\affiliation{Albert Einstein Center for Fundamental Physics, Laboratory for High Energy Physics, University of Bern, Sidlerstrasse 5, CH-3012 Bern, Switzerland}
\affiliation{Department of Physics, Chiba University, 1-33 Yayoi-cho Inage-ku, 263-8522 Chiba, Japan}

\author{Tomoko Ariga\,\orcidlink{0000-0001-9880-3562}}
\affiliation{Kyushu University, 744 Motooka, Nishi-ku, 819-0395 Fukuoka, Japan}

\author{Jeremy Atkinson\,\orcidlink{0009-0003-3287-2196}}
\affiliation{Albert Einstein Center for Fundamental Physics, Laboratory for High Energy Physics, University of Bern, Sidlerstrasse 5, CH-3012 Bern, Switzerland}

\author{Jamie Boyd\,\orcidlink{0000-0001-7360-0726}}
\affiliation{CERN, CH-1211 Geneva 23, Switzerland}

\author{Kohei Chinone\,\orcidlink{0009-0006-5128-7672}}
\affiliation{Department of Physics, Chiba University, 1-33 Yayoi-cho Inage-ku, 263-8522 Chiba, Japan}

\author{Radu Dobre\,\orcidlink{0000-0002-9518-6068}}
\affiliation{Institute of Space Science — INFLPR Subsidiary, 409 Atomiștilor Street, 077125, Bucharest, Romania}

\author{Elena Firu\,\orcidlink{0000-0002-3109-5378}}
\affiliation{Institute of Space Science — INFLPR Subsidiary, 409 Atomiștilor Street, 077125, Bucharest, Romania}

\author{Haruhi Fujimori\,\orcidlink{0009-0002-5026-8497}}
\affiliation{Department of Physics, Chiba University, 1-33 Yayoi-cho Inage-ku, 263-8522 Chiba, Japan}

\author{Stephen Gibson\,\orcidlink{0000-0002-1236-9249}}
\affiliation{Royal Holloway, University of London, Egham, TW20 0EX, United Kingdom}

\author{Daiki Hayakawa\,\orcidlink{0000-0003-4253-4484}}
\affiliation{Department of Physics, Chiba University, 1-33 Yayoi-cho Inage-ku, 263-8522 Chiba, Japan}

\author{Enrique Kajomovitz\,\orcidlink{0000-0002-8464-1790}}
\affiliation{Department of Physics and Astronomy, Technion---Israel Institute of Technology, Haifa 32000, Israel}

\author{Alex Keyken\,\orcidlink{0009-0001-4886-2924}}
\affiliation{Royal Holloway, University of London, Egham, TW20 0EX, United Kingdom}

\author{Umut Kose\,\orcidlink{0000-0001-5380-9354}}
\affiliation{Institute for Particle Physics, ETH Z\"urich, Z\"urich 8093, Switzerland}

\author{Laurie Nevay\,\orcidlink{0000-0001-7225-9327}}
\affiliation{CERN, CH-1211 Geneva 23, Switzerland}

\author{Ken Ohashi\,\orcidlink{0009-0000-9494-8457}}
\affiliation{Albert Einstein Center for Fundamental Physics, Laboratory for High Energy Physics, University of Bern, Sidlerstrasse 5, CH-3012 Bern, Switzerland}
\affiliation{Department of Physics, Chiba University, 1-33 Yayoi-cho Inage-ku, 263-8522 Chiba, Japan}

\author{Simon Thor\,\orcidlink{0000-0002-9183-526X}}
\affiliation{Institute for Particle Physics, ETH Z\"urich, Z\"urich 8093, Switzerland}

%\author{\textcolor{blue}{the FASER$\nu$2 members if they want (not automatic)}}

\begin{abstract}
\bigskip
The Forward Physics Facility (FPF) at the High-Luminosity LHC (HL-LHC) will enable high-statistics measurements of TeV-scale neutrinos, but the intense flux of forward muons poses a major challenge for neutrino detectors in the far-forward region.

We investigate the suppression of background muons using sweeper magnets with a simulation framework combining SIBYLL event generation, BDSIM beam transport, Geant4 particle tracking, and realistic magnetic field maps.

Starting from a muon flux of $3.8\times10^3~\mathrm{cm^{-2}}$ per $\mathrm{fb^{-1}}$ without magnets, a magnet in the LHC tunnel alone achieves the target level of $2\times10^3~\mathrm{cm^{-2}}$ per $\mathrm{fb^{-1}}$. Additional magnets at the TI18 tunnel and FPF entrance further reduce the flux to $1.5\times10^3~\mathrm{cm^{-2}}$ per $\mathrm{fb^{-1}}$ in the optimized configuration.

These results demonstrate that a properly optimized multi-stage sweeper magnet system can significantly reduce the forward muon background, while also highlighting the importance of realistic transport simulations and geometrical constraints in achieving further suppression.

\end{abstract}

\maketitle

%\begin{center}
%\copyright~2026 CERN for the benefit of the FASER Collaboration. Reproduction of this article or parts of it is allowed as specified in the CC-BY-4.0 license.  
%\end{center}

\clearpage
% \tableofcontents
% \clearpage

\section{Introduction}

The forward region of high-energy proton--proton collisions provides a unique environment to study weakly interacting particles at the TeV scale. 
In particular, collider neutrinos produced in the far-forward direction carry valuable information on particle production mechanisms in proton--proton collisions and on neutrino interactions at energies far beyond those accessible in conventional accelerator experiments.
The FASER experiment~\cite{Feng:2017uoz, FASER:2018bac, FASER:2019dxq, FASER:2020gpr, FASER:2022hcn} has demonstrated the feasibility of detecting collider neutrinos at the Large Hadron Collider (LHC)~\cite{FASER:2021mtu, FASER:2023zcr}. It has also performed measurements of neutrino cross sections at TeV energies~\cite{FASER:2024hoe, FASER:2024ref} and measurements of the forward muon flux~\cite{FASER:2019dxq}, opening a new avenue for forward physics measurements~\cite{Ariga:2025qup}.

Looking ahead to the High-Luminosity LHC (HL-LHC) era, the Forward Physics Facility (FPF)~\cite{Anchordoqui:2021ghd, Feng:2022inv, ANCHORDOQUI2026117398} has been proposed to enable a next-generation program with significantly increased detector mass and integrated luminosity. The planned FASER$\nu$2 experiment aims to perform high-precision measurements of all neutrino flavors with a target mass of approximately 20~t, corresponding to orders-of-magnitude larger event statistics than the current detector~\cite{ANCHORDOQUI2026117398}.

\begin{figure}[hb]
    \centering
    \includegraphics[width=0.7\linewidth]{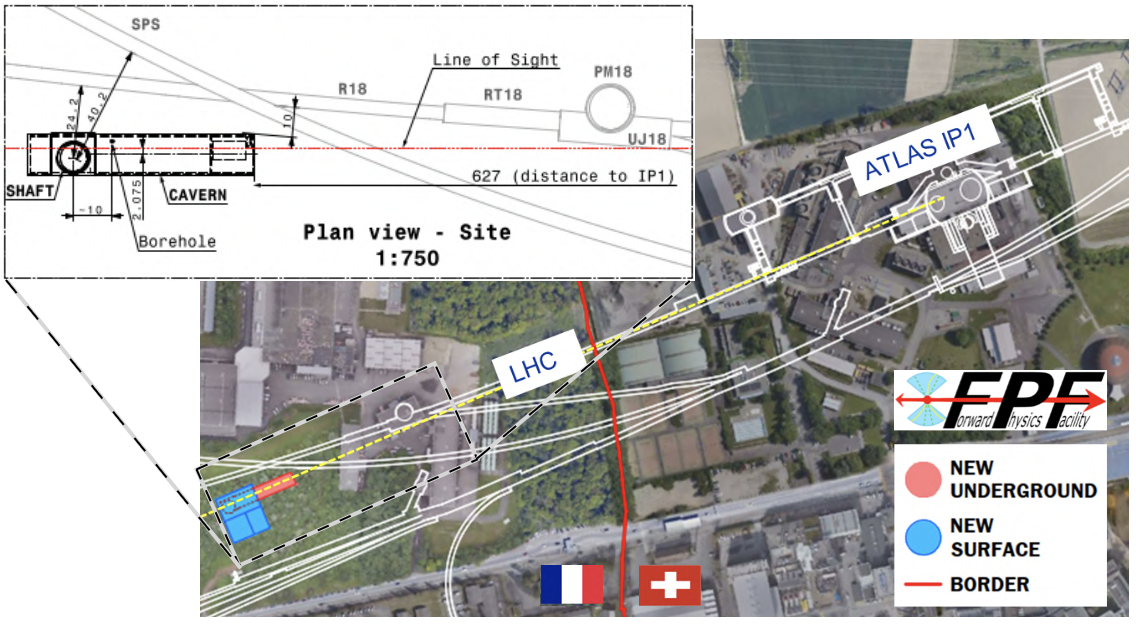}
    \caption{Schematic view of the proposed location of the Forward Physics Facility (FPF)~\cite{ANCHORDOQUI2026117398}. The Line of Sight (LoS) from the interaction point is indicated by the yellow dashed line in the map view and by the red solid line in the inset. The proposed construction areas for the FPF are shown by the red and blue rectangles.}
    \label{fig:fpf_location}
\end{figure}

%A major challenge for such an emulsion-based detector~\cite{ariga2020nuclear} such as FASER$\nu$ and FASER$\nu$2 is the accumulation of background muon tracks. In the current FASER$\nu$~\cite{FASER:2019dxq} operation, the detector must be replaced multiple times per year due to the high density of penetrating muons. For FASER$\nu$2, reducing the replacement frequency to once per year is essential for practical operation. This requires a substantial suppression of the incoming muon flux without affecting the neutrino signal.

A major challenge for such an emulsion-based detector~\cite{ariga2020nuclear} like FASER$\nu$~\cite{FASER:2019dxq} and FASER$\nu$2 is the accumulation of background muon tracks. In the current FASER$\nu$ operation, the detector must be replaced multiple times per year due to the high density of penetrating muons. For FASER$\nu$2, reducing the replacement frequency to approximately once per year is important for practical operation, since the detector will employ about ten times larger film area and approximately 3000 emulsion films, significantly increasing the detector production and replacement cost.

A high track density also degrades the reconstruction performance by increasing the probability of incorrect track matching between emulsion layers. In addition to the primary muon tracks, high-energy muons produce a substantial number of secondary electron tracks through bremsstrahlung and subsequent electromagnetic showers. Since the fraction of these secondary particles increases with muon energy, the tolerable track density depends not only on the muon flux itself but also on the muon energy spectrum. In this study, we target a total track density below approximately $10^6~\mathrm{tracks\ cm^{-2}}$, corresponding to a muon density of about $5\times10^5~\mathrm{cm^{-2}}$. 
This motivates studies of suppressing the incoming muon flux.

Muon backgrounds not only affect track reconstruction in emulsion detectors but also lead to charge pile-up in the liquid-argon detectors (FLArE~\cite{ANCHORDOQUI2026117398}), resulting in space-charge effects that distort the drift electric field and degrade reconstruction performance.

A promising approach is to install a magnetic system upstream of the detector to deflect charged particles away from the detector acceptance. Since neutrinos are electrically neutral, they are unaffected by the magnetic field, while muons can be efficiently displaced depending on the magnetic field configuration. Previous studies have evaluated the forward muon flux using particle transport simulations based on BDSIM~\cite{bdsim_paper}, which have been validated against FASER measurements. In addition, exploratory studies by the CERN STI group using FLUKA~\cite{Battistoni:2015epi, LHCFLUKAmodel} investigated the use of magnetic fields for muon suppression using detailed magnetic field models~\cite{PBCnote}. These studies have highlighted the importance of multiple Coulomb scattering in the surrounding material, which can partially refill the detector acceptance even after magnetic deflection.
% \ken{Citation for FLUKA might be: \cite{Battistoni:2015epi, LHCFLUKAmodel}.} --> Thank you!

Building on this understanding, the present study investigates magnet configurations that enhance the transverse coverage and optimize the placement of sweeper magnets, thereby mitigating the re-entry of muons into the detector acceptance caused by multiple scattering.

We perform a comprehensive and realistic evaluation of muon background suppression using a full simulation chain combining proton--proton collision event generation, beamline transport with BDSIM, and particle tracking with Geant4~\cite{Geant4:2003}. Detailed magnetic field maps are calculated using finite element methods and implemented in the tracking simulation. 
This framework enables a systematic assessment of various magnet configurations and candidate installation locations in the downstream region.

The goal of this study is to identify feasible magnet configurations that achieve the target muon flux for optimal FASER$\nu$2 operation and performance, while satisfying the constraints of the LHC infrastructure. The results provide important input for the design of the FPF and future forward neutrino experiments.

\section{Simulation framework}

A comprehensive simulation framework is developed to evaluate the suppression of background muons by a sweeper magnet system. The simulation consists of a multi-stage chain combining proton--proton collision event generation, beamline transport, particle tracking, and magnetic field modeling.

The initial particle distributions are generated using the SIBYLL 2.3d hadronic interaction model~\cite{Riehn:2019jet} with the CRMC simulation package~\cite{ulrich_2021_5270381} for proton--proton collisions at $\sqrt{s}=14$~TeV.
%The initial particle distributions are generated using the SIBYLL 2.3d hadronic interaction model ~\cite{Riehn:2019jet} with the CRMC simulation package~\cite{ulrich_2021_5270381}. 
These particles are transported through the LHC beamline using BDSIM, which provides realistic distributions of particles reaching the forward region. The output of BDSIM is then used as input to a dedicated Geant4 simulation, where particle propagation through the downstream tunnel geometry and magnetic field configurations is modeled in detail.
%\ken{If Alex uses CRMC simulation package, it is better to cite it. CRMC : Reference : C. Baus, T. Pierog and R. Ulrich. See also: https://doi.org/10.5281/zenodo.4558705}

Magnetic field maps are calculated using finite element analysis and implemented in the Geant4 simulation. This approach allows for a realistic description of the field configuration and its impact on charged particle trajectories. The resulting particle distributions at the detector location are used to evaluate the muon flux and the effectiveness of different magnet configurations.

\subsection{Input muon sample}

%\aki{Alex, please modify as you with. If possible add the passes of muons reaching the FASERnu2 detector.}
%\alex{I couldn't find the TI12 plot, if I have time I will make one for the FPF by monday.}

The input muon sample is obtained from BDSIM simulations of proton--proton collisions at the LHC. BDSIM is a Geant4-based toolkit for particle transport in accelerator environments, including beam optics, magnetic elements, and interactions with surrounding materials. 
The BDSIM framework used in this study has been compared with muon flux measurements at the FASER detector, showing agreement at the 20\% level~\cite{FASER:2025muonflux}. While some discrepancies remain in the detailed modeling of muon transport, the framework provides a reasonable description of the forward muon distributions for the purpose of this study.
%The simulation framework used in this study has been validated against muon flux measurements at the FASER detector, providing confidence in the predicted forward muon distributions.

The muon distributions are extracted at a position approximately 370~m downstream of the interaction point, corresponding to the location where the Line of Sight (LoS) exits the beam pipe and enters the tunnel environment. These distributions are used as initial conditions for the downstream Geant4 simulation.

In the absence of a sweeper magnet, these muons propagate downstream and constitute the baseline muon flux of $3.76\times10^3~\mathrm{cm^{-2}}/\mathrm{fb}^{-1}$ at the FASER$\nu$2 detector location, evaluated within the detector acceptance of $25~\mathrm{cm} \times 64~\mathrm{cm}$.

\begin{figure}[t]
    \centering
    \includegraphics[width=0.34\linewidth]{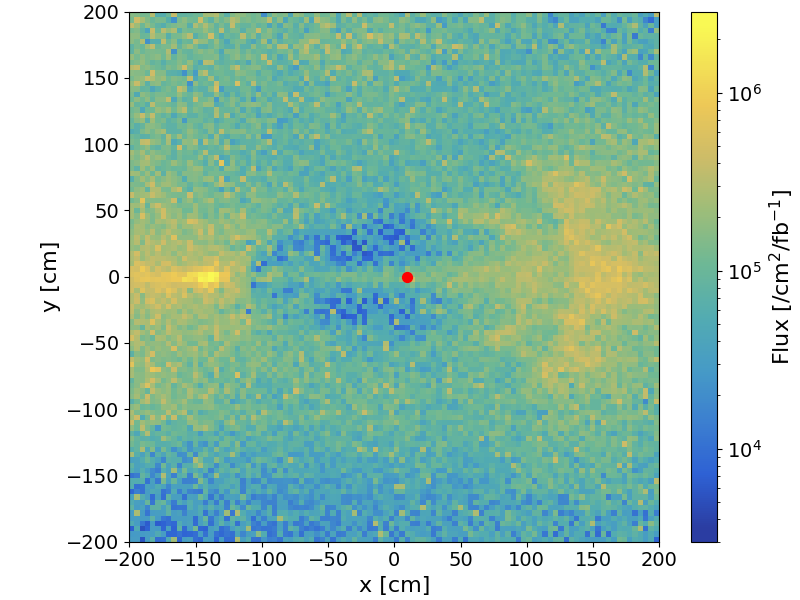}
    \includegraphics[width=0.30\linewidth]{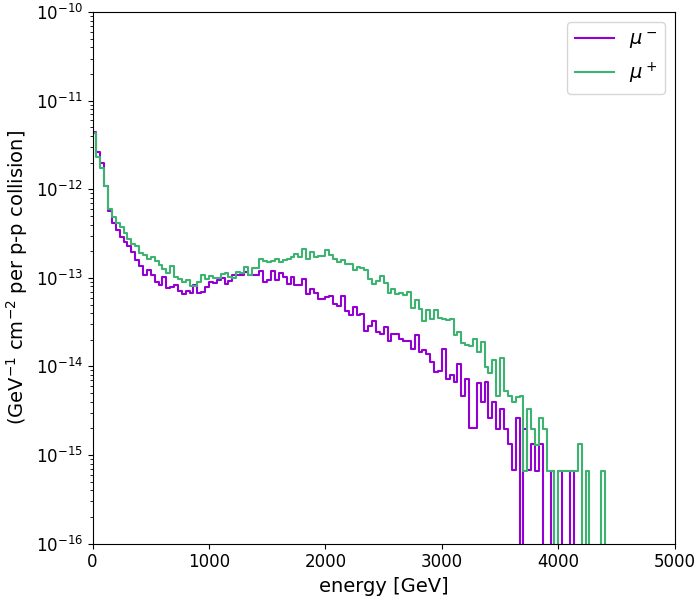}
    \includegraphics[width=0.34\linewidth]{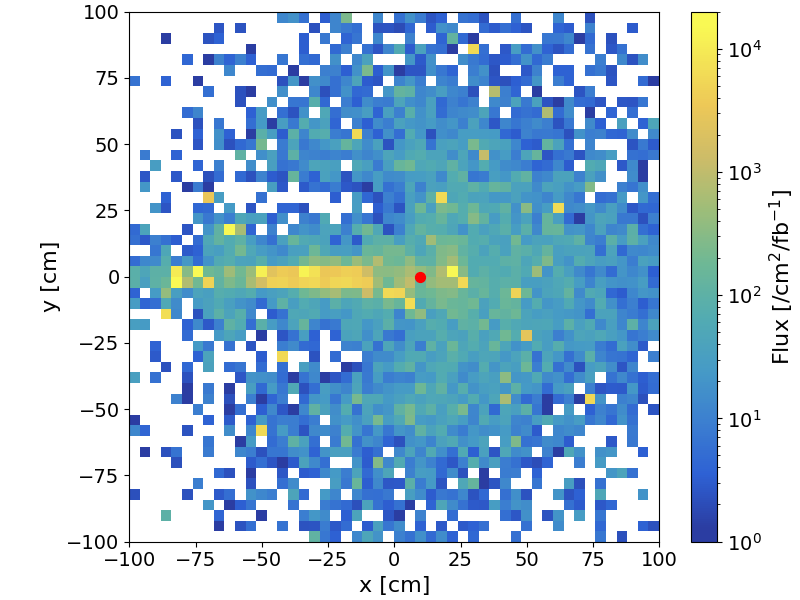}
    \caption{
    Input muon distributions obtained from the BDSIM simulation at a position 370~m downstream of the proton--proton interaction point. Left: transverse position distribution of muons, where the red marker indicates the position of the LoS. The distribution includes both $\mu^-$ and $\mu^+$ and reflects the deflection caused by the LHC magnetic lattice. Middle: energy distributions of the input muons, where $\mu^-$ and $\mu^+$ are shown in purple and green, respectively. Right: transverse distribution at 370~m of muons that reach the FASER$\nu$2 detector location in the absence of a sweeper magnet, evaluated within the detector acceptance of $25~\mathrm{cm} \times 64~\mathrm{cm}$.
    }
    \label{fig:inputmuons}
\end{figure}

Figure~\ref{fig:inputmuons} shows the transverse position distribution of the input muons, the energy distributions of $\mu^\pm$, and the transverse distribution of muons reaching the FASER$\nu$2 detector location. A large fraction of the muons have energies below 200~GeV, while a broad secondary peak is observed above 1~TeV. The muon distribution for those reaching the FASER$\nu$2 location exhibits a strong asymmetry with respect to the LoS, with a concentration of muons on the side closer to the LHC beam pipe. This feature reflects the deflection of charged particles by the LHC magnetic lattice during their propagation downstream from the interaction point.

These characteristics play an important role in the optimization of the sweeper magnet configuration, since the effectiveness of magnetic deflection depends strongly on the initial spatial and angular distributions of the muons.

\subsection{Muon transport simulation}

The transport of muons from the input plane at 370~m to the FASER$\nu$2 detector location is simulated using a dedicated Geant4-based setup. The simulation is designed to provide a realistic estimate of the sweeper magnet performance while keeping the computational cost manageable.

A simplified geometrical model is implemented. The coordinate system is defined as a right-handed system with the $z$-axis aligned along the LHC beam direction. 
The region downstream of the 370~m point is modeled with the essential elements relevant for muon propagation. The surrounding rock is implemented as a uniform volume of silicon dioxide, extending sufficiently in the transverse direction to account for Multiple Coulomb Scattering (MCS) effects. The LHC tunnel is modeled as an air-filled cylindrical volume, followed by the intermediate TI18 tunnel and the FPF region, each represented by simplified cylindrical geometries with dimensions corresponding to the actual layout.

To evaluate the muon transport and the effect of the sweeper magnets, particle scoring planes are placed at four locations: the input plane (370~m), the TI18 region ($\sim$480~m), the FPF entrance, and the FASER$\nu$2 detector position. At each plane, the particle position, direction, and energy are recorded.

The simulation focuses on muon propagation and magnetic deflection; therefore, electron tracking is disabled to reduce the computational load. Accelerator components such as beam pipes and LHC magnets are not explicitly included, as their impact on muon transport in this downstream region is subdominant for high momentum particles.
%compared to scattering in the surrounding material.
%\alex{Maybe clarify that the impact is subdominant for highly collimated / high E particles.}

\subsection{Biasing and resampling}

The production of high-energy muons in the far-forward region is a rare process, with typically one muon reaching the detector per $\mathcal{O}(10^6$--$10^7)$ proton--proton collisions. To obtain sufficient statistical precision, biasing techniques are applied at the event generation and transport stages.

Cross-section biasing is used to enhance processes that contribute to muon production, including $\pi^\pm$ and $K^\pm$ decays, as well as electromagnetic processes such as $e^+e^-$ annihilation. The bias factors are applied in the BDSIM simulation and are accounted for by assigning appropriate statistical weights to each event.

In addition, a resampling procedure based on splitting and \textit{Russian roulette} is applied to the muon sample. In the splitting step, muons are replicated multiple times and propagated independently, while their statistical weights are reduced accordingly. Conversely, Russian roulette removes low-weight particles probabilistically, preserving the total weight on average. These techniques reduce the variance of event weights and improve the effective statistical power of the simulation.

The consistency of the weighted simulation is verified by ensuring that the total weight is conserved and that the resulting muon flux agrees with reference simulations within statistical uncertainties. The resampling procedure increases the effective statistics by several orders of magnitude, significantly improving the precision of the flux estimation.
%\alex{Several orders (up to $10^5$) of magnitude (unless Russian Roulette reduces it significantly).}

These simulated muon distributions are used as inputs to evaluate the effectiveness of various magnet configurations, as described in the following section.

\section{Magnet system}

\subsection{Conceptual study of muon deflection and scattering}

To gain qualitative insight into the interplay between magnetic deflection and multiple scattering, a simplified toy Monte Carlo study was performed. Muons were generated with a uniform distribution in the transverse ($x$) direction and propagated along the $z$-axis. In this study, only magnetic deflection and multiple Coulomb scattering in rock was considered.

\begin{figure}[t]
    \centering
    \begin{minipage}{0.32\textwidth}
        \centering
        \includegraphics[width=\linewidth]{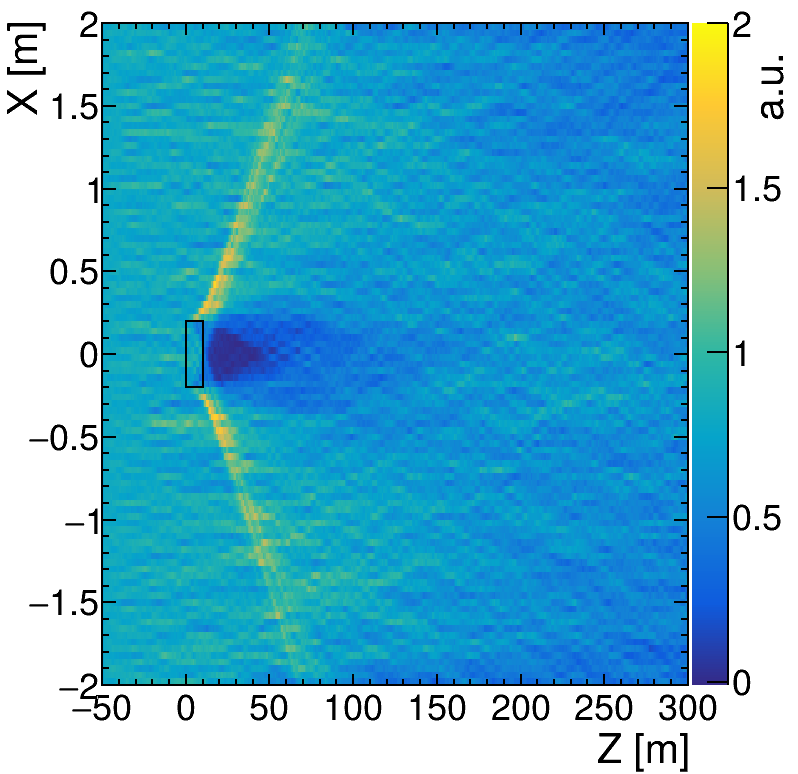}
        \par (a) 100~GeV
    \end{minipage}\hfill
    \begin{minipage}{0.32\textwidth}
        \centering
        \includegraphics[width=\linewidth]{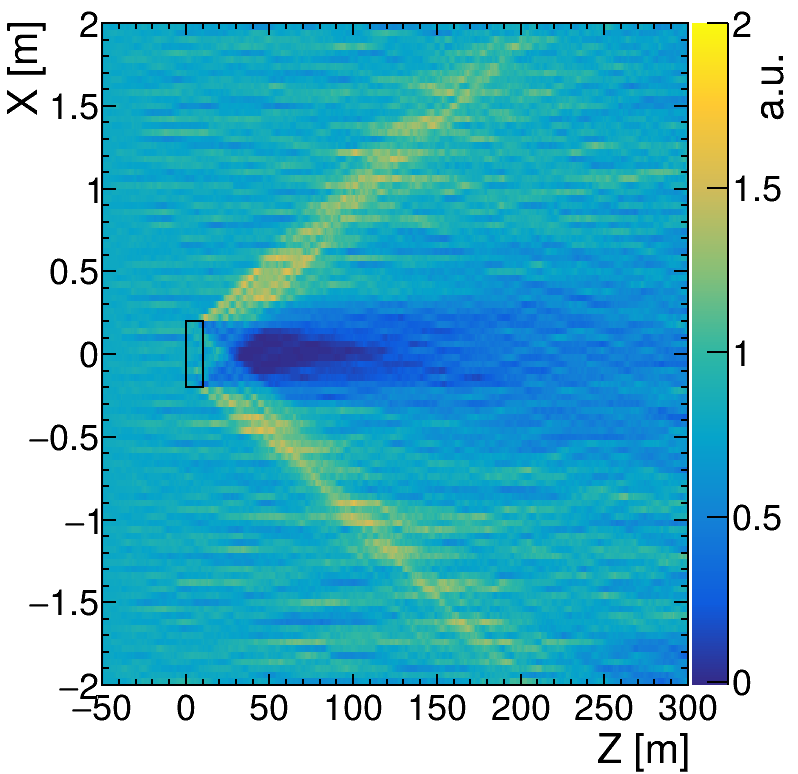}
        \par (b) 300~GeV
    \end{minipage}\hfill
    \begin{minipage}{0.32\textwidth}
        \centering
        \includegraphics[width=\linewidth]{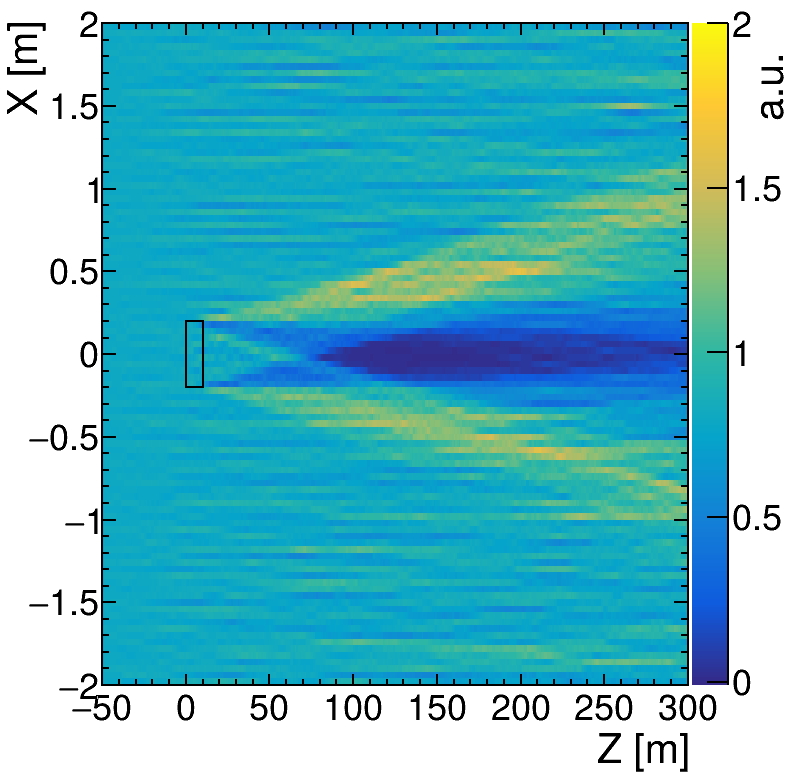}
        \par (c) 1~TeV
    \end{minipage}
\caption{
Muon density distributions in the $x$--$z$ plane obtained from a toy Monte Carlo simulation for (a) 100~GeV, (b) 300~GeV, and (c) 1~TeV muons. Muons are injected in the $z$ direction. A 1~T magnetic field region with a transverse width of $\pm 20$~cm in $x$ and a longitudinal length of 5~m is placed at $(x,z) = (0,0)$. The color scale represents the muon density normalized to the incident distribution. Only magnetic deflection and multiple Coulomb scattering in rock are considered in this toy simulation. The muon charge is equally randomized among $\mu^+$ and $\mu^-$.
}
\label{fig:toymc}
\end{figure}

Figure~\ref{fig:toymc} shows the evolution of the muon density in the $x$--$z$ plane for different muon energies. Immediately downstream of the magnet, a clear reduction of the muon density near the LoS is observed, indicating effective deflection of low-energy components. However, as the muons propagate further, multiple Coulomb scattering broadens their trajectories, and a fraction of the muons re-enters the detector acceptance.

This effect depends strongly on the muon energy. Lower-energy muons are more efficiently deflected but are also more susceptible to scattering, while higher-energy muons are less affected by both processes. As a result, the balance between magnetic deflection and multiple scattering leads to an energy-dependent optimal distance between the magnet and the detector.

%This behavior provides a possible explanation for the limited suppression observed in previous studies using simplified magnetic field configurations, particularly for magnets placed along the LoS.

This behavior provides a possible explanation for the limited suppression observed in previous studies using simplified magnetic field configurations, particularly for magnets placed along the LoS. In fact, previous studies~\cite{PBCnote} have already pointed out that multiple Coulomb scattering in the rock downstream of the sweeper magnet can partially refill low-fluence regions after magnetic deflection. The present study builds on this understanding by investigating magnet configurations with improved transverse coverage and optimized placement to mitigate the re-entry of scattered muons into the detector acceptance.

\subsection{Magnet placement}

Three candidate locations for the installation of a sweeper magnet are considered in this study: the LHC tunnel, the TI18 tunnel, and the entrance of the FPF. These locations correspond to different stages of particle propagation from the interaction point and offer distinct geometrical and operational constraints.

The LHC tunnel location, situated approximately 370~m downstream of the ATLAS interaction point where the LoS exits the beam pipe, provides the earliest opportunity for muon deflection before further downstream propagation. However, the available space is constrained by existing LHC infrastructure, requiring the sweeper magnets to be installed within the narrow region between the tunnel wall and the helium transfer line.

The TI18 tunnel, located around 480~m downstream of the interaction point, provides an intermediate location within the existing tunnel infrastructure. This region is shared with the SND@HL-LHC detector~\cite{Abbaneo:2926288}, resulting in limited flexibility for magnet placement.

%The entrance of the FPF, at approximately 627~m from the interaction point, is closest to the detector location. While installation at this position is subject to space constraints due to the presence of multiple experiments, it provides the shortest lever arm for deflected muons to move away from the LoS, allowing direct control of the particle flux entering the detector volume. This configuration is therefore particularly effective for suppressing low-energy muons with larger magnetic deflection.

%The entrance of the FPF, at approximately 627~m from the interaction point, is closest to the detector location. While installation at this position is subject to space constraints due to the presence of multiple experiments, it provides the shortest lever arm for deflected muons to move away from the LoS, allowing direct control of the particle flux entering the detector volume and being particularly effective for suppressing low-energy muons with larger magnetic deflection.

The entrance of the FPF, at approximately 627~m from the interaction point, is closest to the detector location. While installation at this position is subject to space constraints due to the presence of multiple experiments, it allows direct control of the particle flux entering the detector volume. However, the short distance to the detector limits the lever arm for deflected muons to move away from the LoS, making this configuration primarily effective for suppressing low-energy muons with larger magnetic deflection.
%\alex{From these three repeated paragraphs I prefer the 3rd, emphasizing the FPF location is favorable for the deflection of lower energy muons.}

\begin{figure}[t]
\centering
\includegraphics[width=0.8\textwidth]{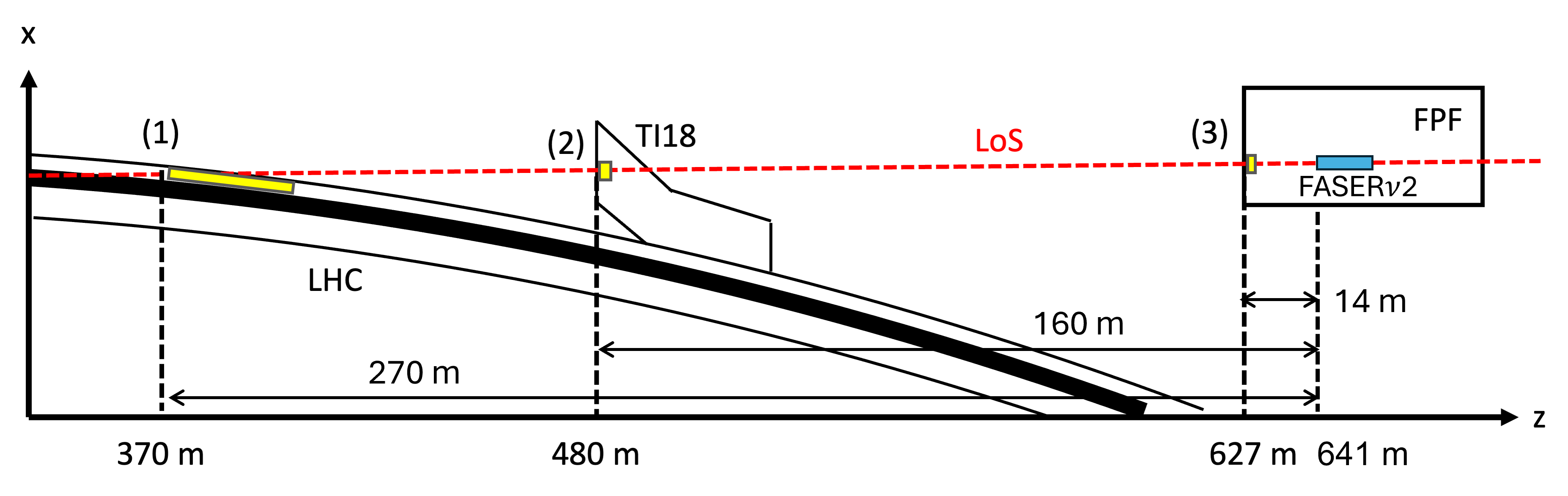}
\caption{
Schematic view of the LHC beamline and the candidate locations for the sweeper magnet installation. The positions correspond to the (1) LHC tunnel ($\sim$370~m), (2) TI18 tunnel ($\sim$480~m), and (3) the FPF entrance ($\sim$627~m) downstream of the interaction point. 
}
\label{fig:magnet_locations}
\end{figure}

A schematic overview of the considered magnet locations is shown in Fig.~\ref{fig:magnet_locations}.

\subsection{Magnet configuration}

The sweeper magnet is designed to deflect charged particles away from the detector acceptance while leaving neutrinos unaffected. The effectiveness of the deflection depends on the magnetic field strength, the magnet length, and its position relative to the LoS.

For the LHC tunnel installation candidate (location (1)), the high-radiation environment makes it difficult to ensure stable operation of power converters. For this reason, a configuration based on permanent samarium--cobalt (SmCo) magnets combined with iron yokes for magnetic flux return is considered instead of superconducting magnets. SmCo magnets are known to have higher radiation tolerance than neodymium magnets and are therefore expected to maintain stable magnetic performance in the radiation environment of the LHC tunnel. In this study, a magnetic field strength of 1~T is assumed for the SmCo magnet system.

\begin{figure}[t]
    \centering

    \begin{tabular}{cccccc}
    \includegraphics[width=0.13\linewidth]{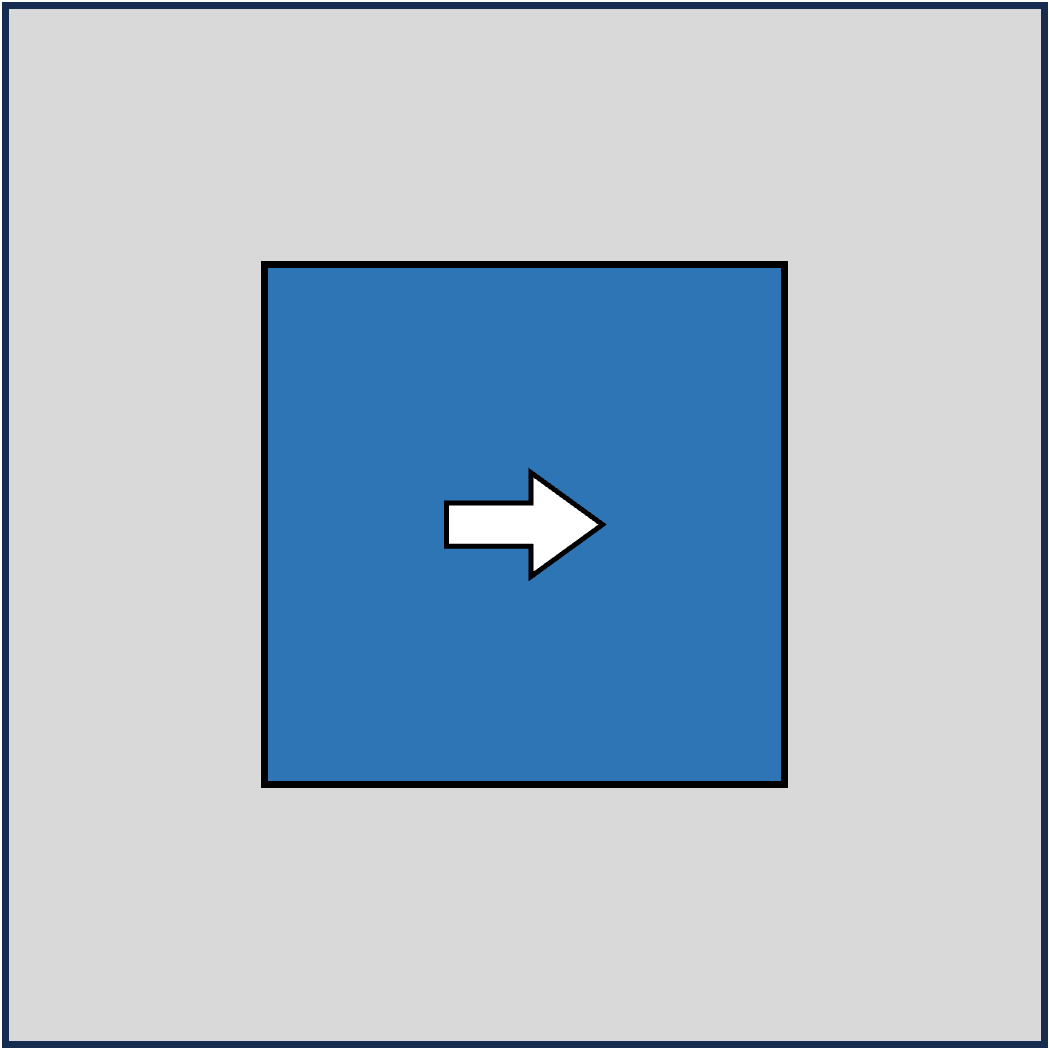} &
    \includegraphics[width=0.13\linewidth]{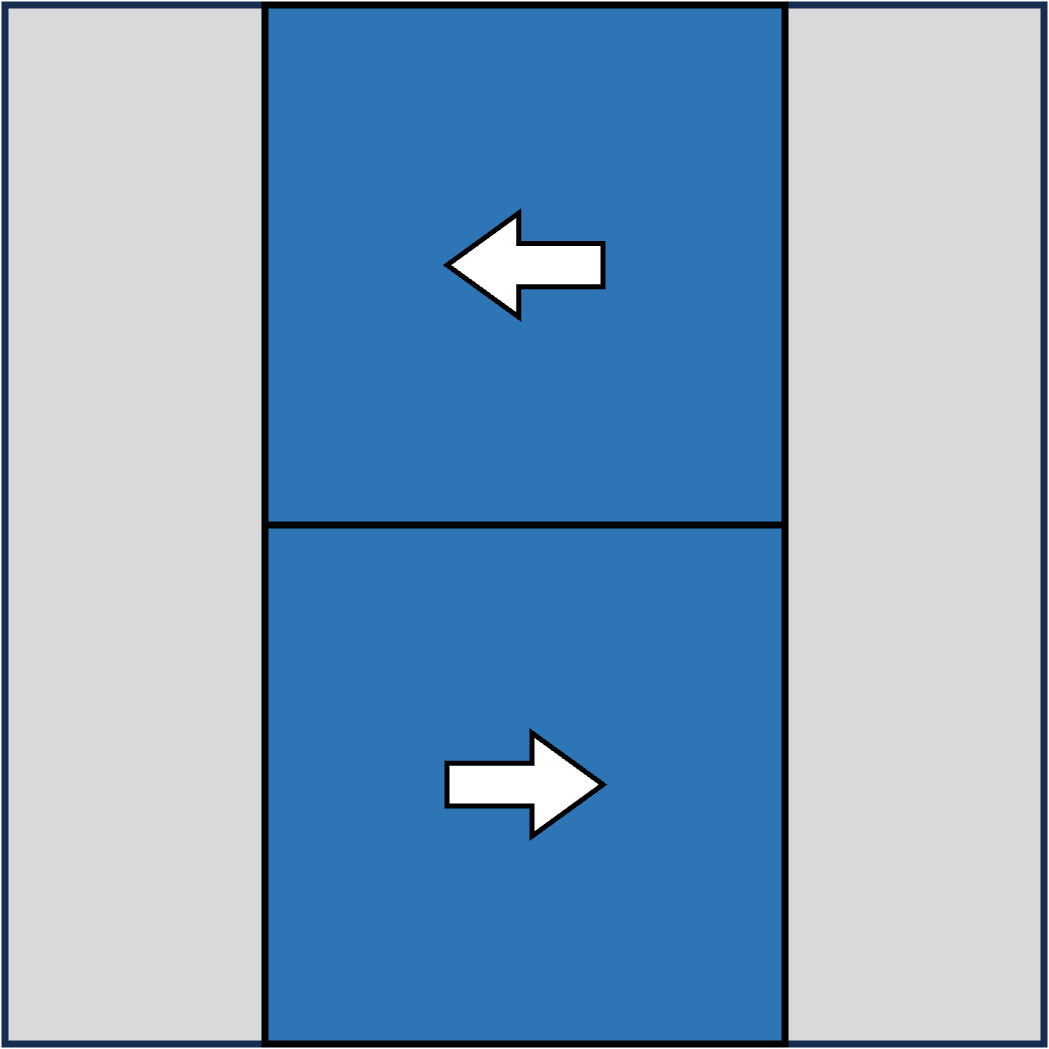} &
    \includegraphics[width=0.13\linewidth]{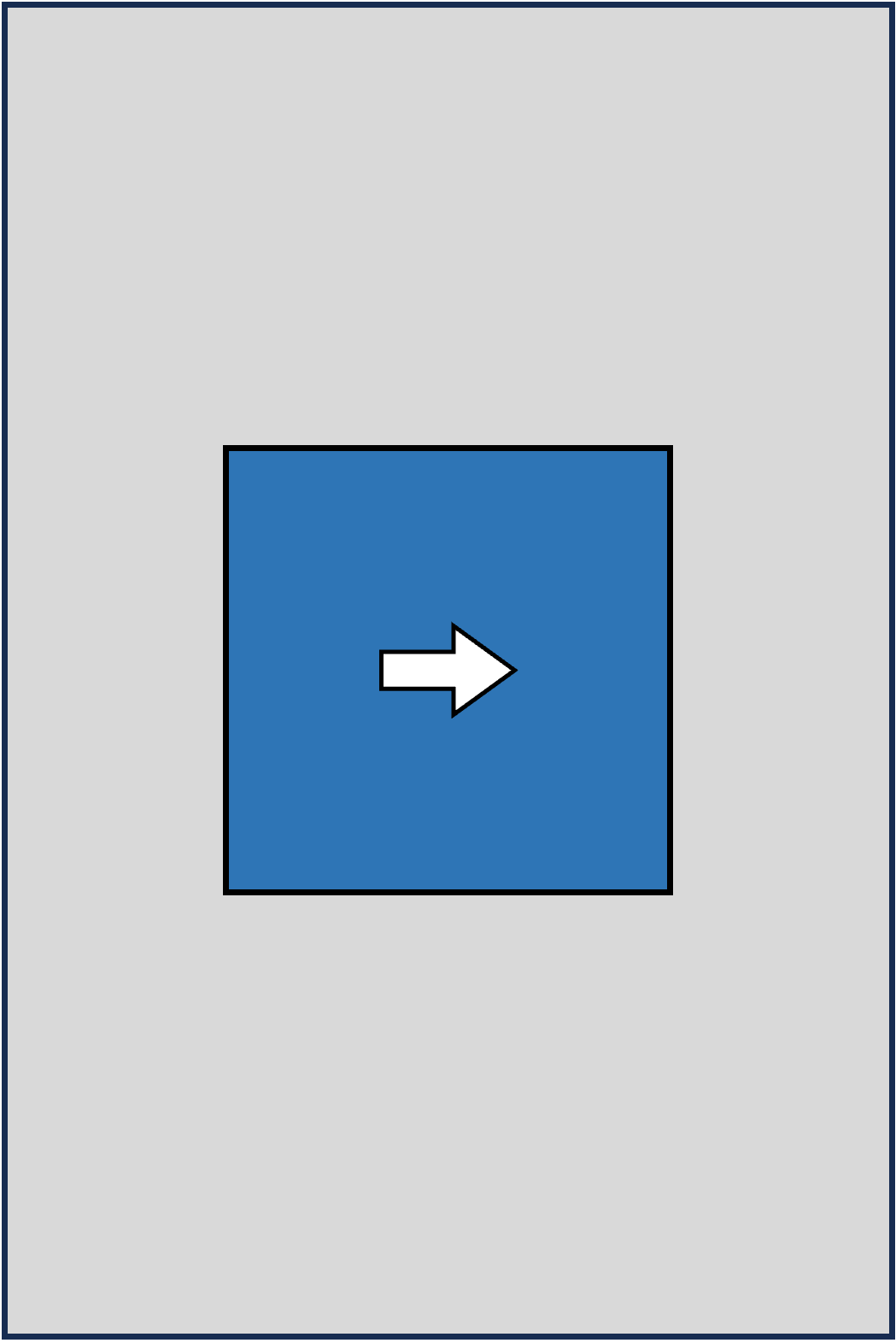} &
    \includegraphics[width=0.13\linewidth]{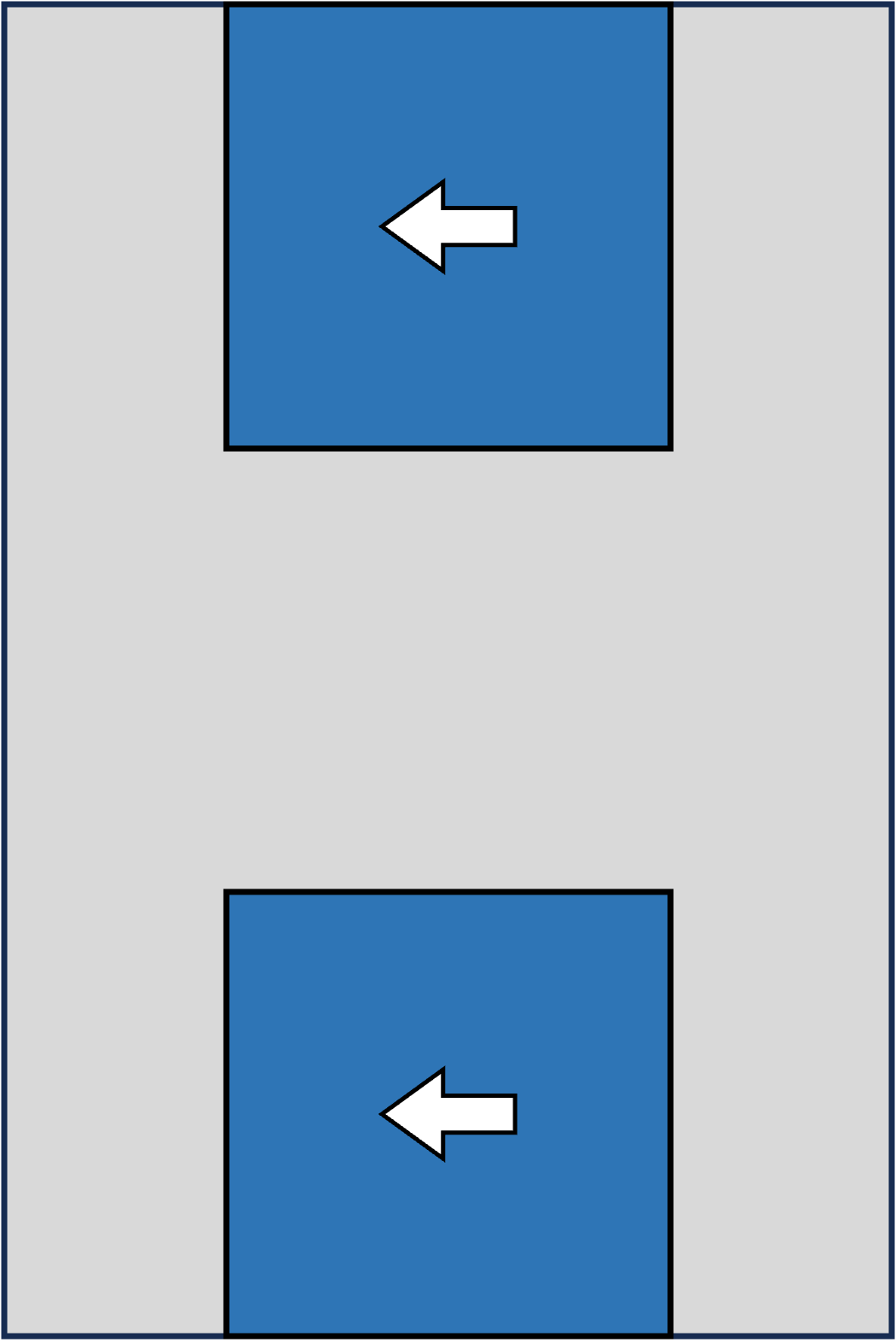} &
    \includegraphics[width=0.2\linewidth]{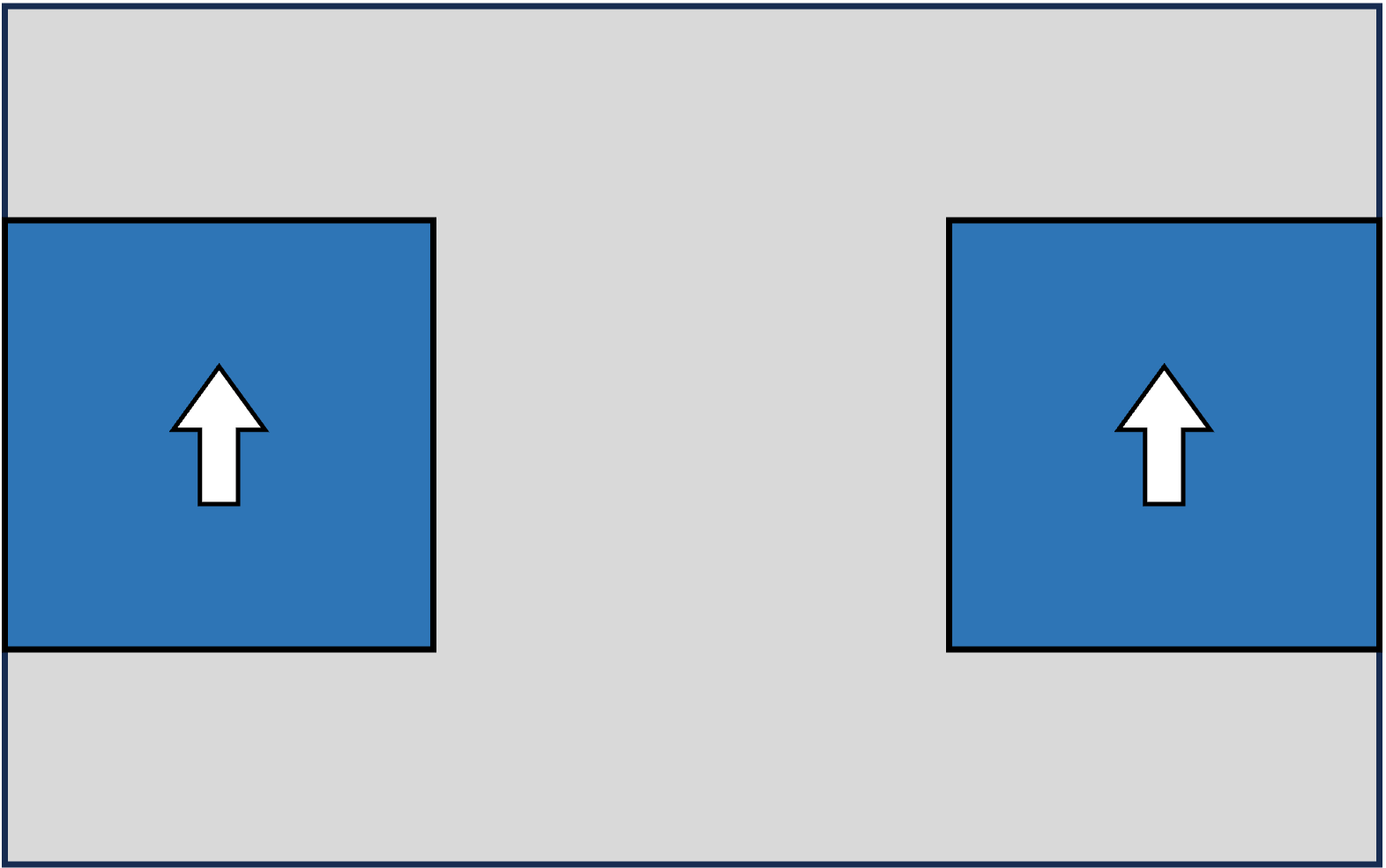} &
    \includegraphics[width=0.1\linewidth]{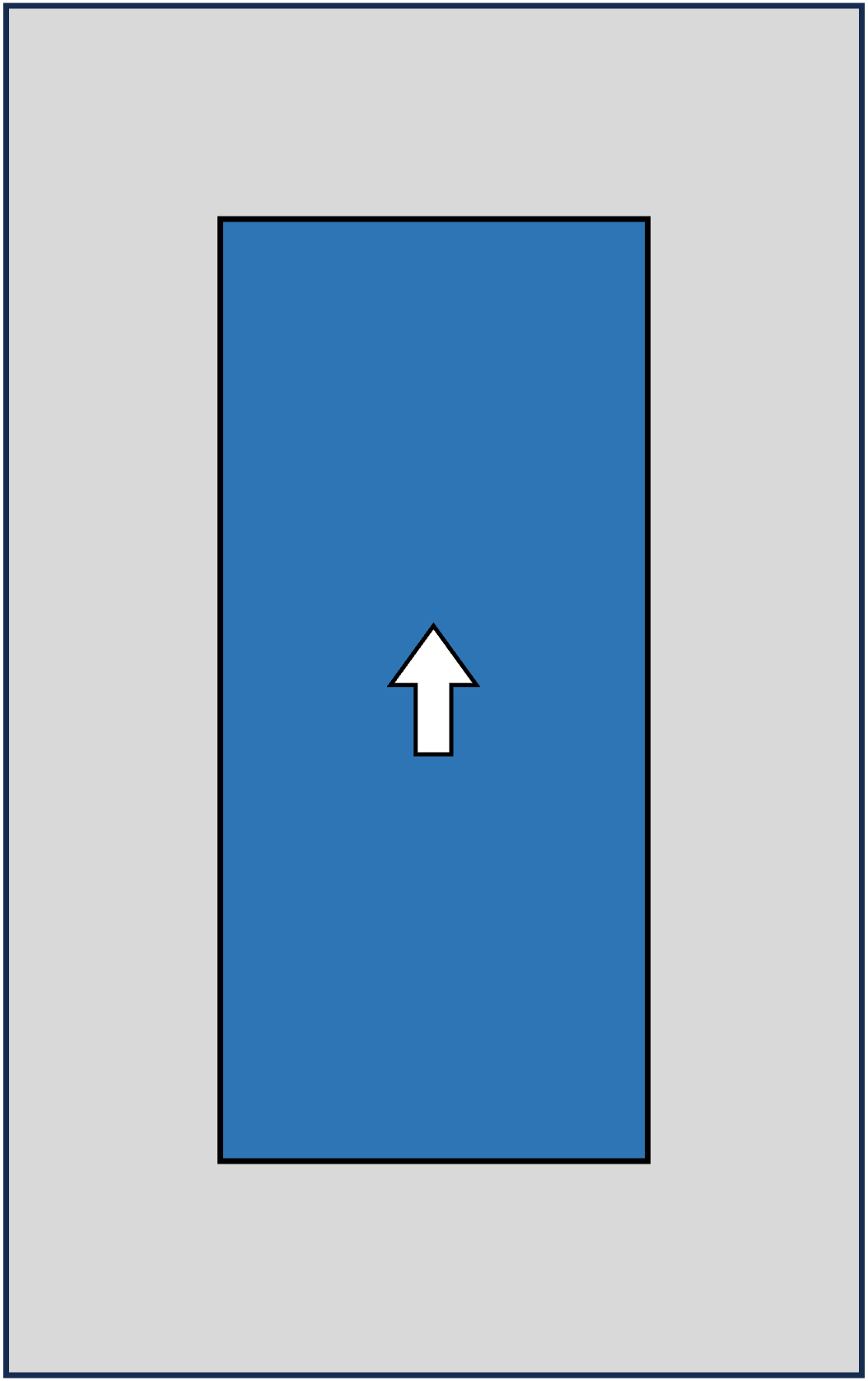} \\
    LHC-A, TI18-A & LHC-B & LHC-C & LHC-D & TI18-B & FPF-A
    \end{tabular}

    \caption{
    Cross sections of the magnet configurations considered for each installation location. The outer dimensions are $40~\mathrm{cm} \times 40~\mathrm{cm}$ for LHC-A, LHC-B, and TI18-A; $40~\mathrm{cm} \times 80~\mathrm{cm}$ for LHC-C and LHC-D; $80~\mathrm{cm} \times 50~\mathrm{cm}$ for TI18-B; and $25~\mathrm{cm} \times 64~\mathrm{cm}$ for FPF-A.
    }
    \label{fig:magnet_crosssections}
\end{figure}

The magnet configuration is adapted for each candidate installation location, taking into account geometrical constraints and the spatial profile of the background muons. The cross sections of the magnet configurations considered in this study are shown in Fig.~\ref{fig:magnet_crosssections}.

For the LHC tunnel location, a long tunnel-parallel magnet configuration is considered. The configuration is characterized by the transverse offset from the LoS, $\Delta x$, and the magnet length along the beam direction, $\ell$. Here, $\Delta x$ is defined as the horizontal displacement of the magnet center relative to the LoS.

To enhance the deflection efficiency for the asymmetric muon distribution shown in Fig.~\ref{fig:inputmuons}, the magnet is positioned close to the LoS with a finite offset. The magnet length $\ell$ is varied between 18~m and 36~m to evaluate its impact on the suppression performance. Three transverse magnet sizes are considered: $20~\mathrm{cm} \times 20~\mathrm{cm}$, $40~\mathrm{cm} \times 40~\mathrm{cm}$ and $40~\mathrm{cm} \times 80~\mathrm{cm}$.

In the TI18 tunnel, the available space imposes additional constraints on the magnet dimensions and placement due to the presence of the SND@LHC detector~\cite{Abbaneo:2926288}. In this region, the magnet length is limited to 1~m.

At the entrance of the FPF, the available space along the beam direction is strongly constrained by the presence of multiple detectors~\cite{ANCHORDOQUI2026117398}. Therefore, compact magnet configurations with a length of 1~m are considered. Since this location is only 14~m upstream of the FASER$\nu$2 detector, the transverse dimensions of the magnet are chosen to match the detector dimensions of $25~\mathrm{cm} \times 64~\mathrm{cm}$.

For each location, multiple configurations are evaluated by scanning over the transverse offset $\Delta x$ and the magnet length $\ell$. The resulting muon flux at the detector position is used to quantify the suppression performance.

\subsection{Magnetic field modeling}

To evaluate realistic magnetic deflection, the magnetic field distributions of the proposed magnet configurations were calculated using the finite-element analysis package Elmer~\cite{Elmer}. For each magnet configuration, a two-dimensional static magnetic-field calculation was performed including the permanent-magnet blocks and iron yokes.

\begin{figure}[t]
    \centering
    \begin{tabular}{ccc}
    \includegraphics[width=0.30\linewidth, trim={1cm 0 0 1cm}, clip]{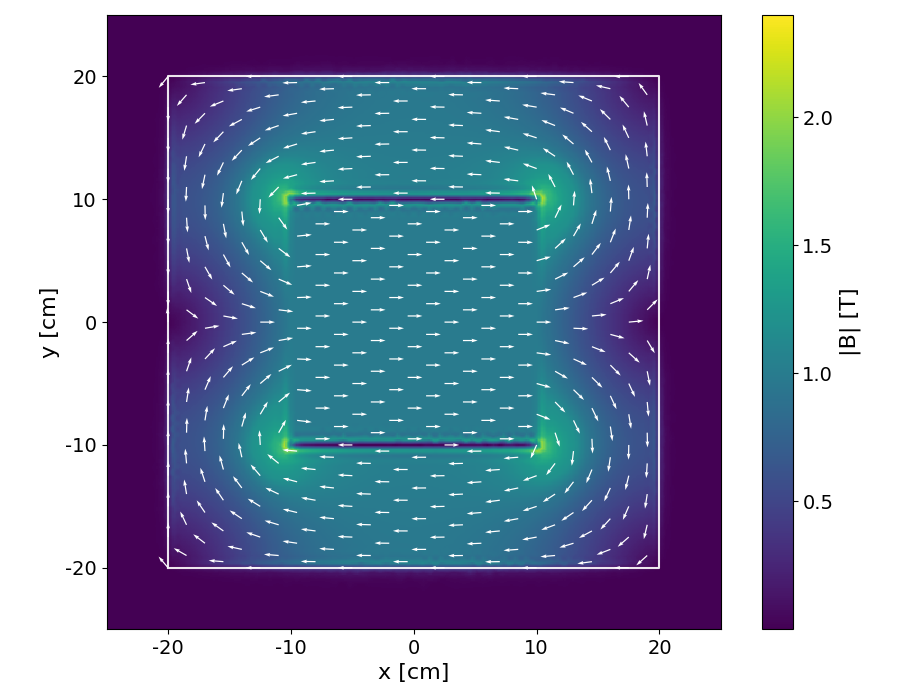} &
    \includegraphics[width=0.33\linewidth, trim={5.2cm 0 0 1cm}, clip]{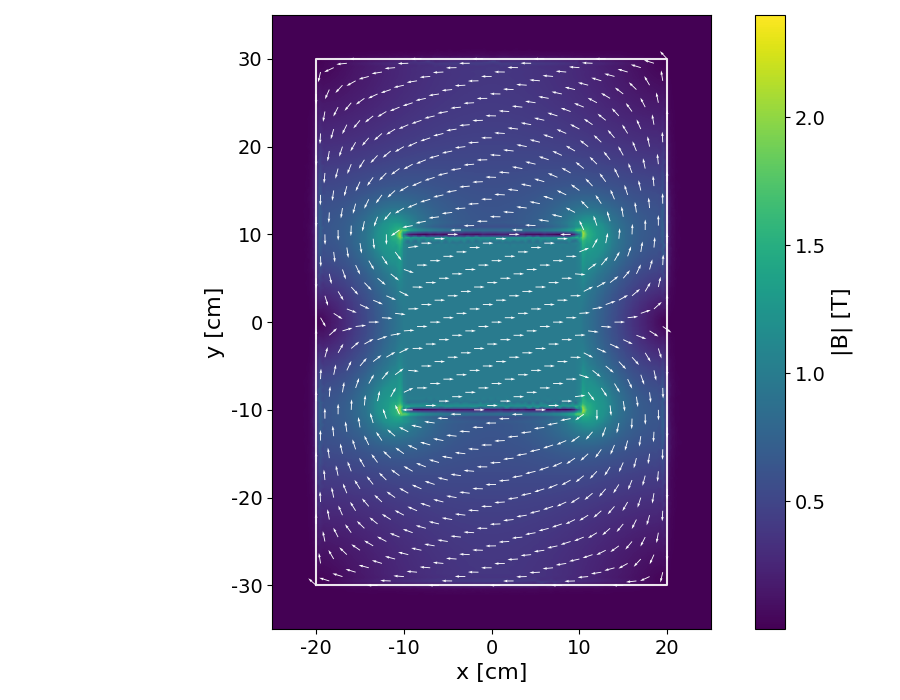} &
    \includegraphics[width=0.33\linewidth, trim={5.2cm 0 0 1cm}, clip]{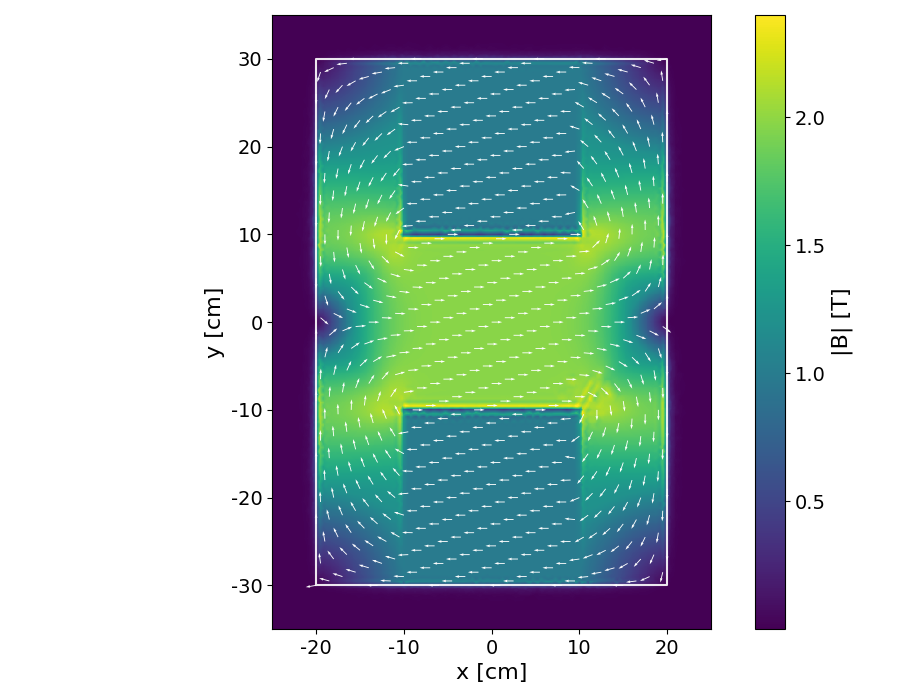} \\
    LHC-A & LHC-C & LHC-D
    \end{tabular}

    \caption{
    Magnetic flux density distributions obtained from two-dimensional finite-element calculations with Elmer for the LHC-A, LHC-C, and LHC-D magnet configurations. The color scale indicates the magnetic flux density, while the arrows represent the magnetic field direction.
    }
    \label{fig:elmer}
\end{figure}

Figure~\ref{fig:elmer} shows examples of the magnetic flux density distributions for the LHC-A, LHC-C, and LHC-D configurations. In the LHC-A and LHC-C configurations, the magnetic field strength in the central region is approximately 1~T. In contrast, the LHC-D configuration produces magnetic fields in the yoke region that are approximately twice as strong due to the coupled magnetic structure.

In the LHC-C configuration, the magnetic flux density in the outer yoke regions ($y < -10$~cm and $y > 10$~cm) is intentionally reduced in order to avoid focusing background muons toward the detector acceptance.

The resulting magnetic field maps were extruded uniformly along the longitudinal direction and implemented in the Geant4 simulation for particle tracking.

\section{Evaluation metrics}
\label{sec:evaluation}

The performance of the sweeper magnet configurations is evaluated at the upstream surface of the FASER$\nu$2 detector ($z = 641$~m). Unless otherwise noted, the evaluation region corresponds to the current detector geometry of $25~\mathrm{cm} \times 64~\mathrm{cm}$.

For each configuration, the weighted muon yield in the evaluation region is calculated as $N = \sum_i w_i$, where $w_i$ is the statistical weight of the $i$-th event. Statistical uncertainties are evaluated using the standard treatment for weighted Poisson event samples.

The suppression performance is quantified by the residual ratio

\begin{equation}
R =
\frac{N_{\mathrm{with\ magnet}}}
     {N_{\mathrm{without\ magnet}}},
\end{equation}

where smaller values of $R$ correspond to stronger muon suppression.

The muon flux is defined as

\begin{equation}
\Phi
=
\frac{1}{S}
\left(
\sum_i w_i
\right)
\frac{10^{14}}
     {N_{\mathrm{p-p}}}
\quad
\mathrm{cm^{-2}}/\mathrm{fb}^{-1},
\end{equation}

where $S$ is the evaluation area and
$N_{\mathrm{p-p}}$ is the number of simulated proton--proton collisions.

Assuming an annual integrated luminosity of 250~fb$^{-1}$ at the HL-LHC and a maximum tolerable track density of $5\times10^5~\mathrm{cm^{-2}}$, the target muon flux for annual detector replacement is

\begin{equation}
\Phi_{\mathrm{target}}
=
2.0 \times 10^3~
\mathrm{cm^{-2}}/\mathrm{fb}^{-1}.
\end{equation}

The magnet configurations are evaluated using both the residual ratio $R$ and the resulting muon flux $\Phi$ relative to this target value.

\section{Results}

\subsection{Muon suppression with magnets in the LHC tunnel}

We first evaluate the suppression of background muons using magnets installed in the LHC tunnel. Four magnet configurations, denoted LHC-A to LHC-D, are studied by varying the magnet length $\ell$ and transverse offset $\Delta x$.

\begin{figure}
    \centering
        \centering
        \includegraphics[width=0.28\linewidth]{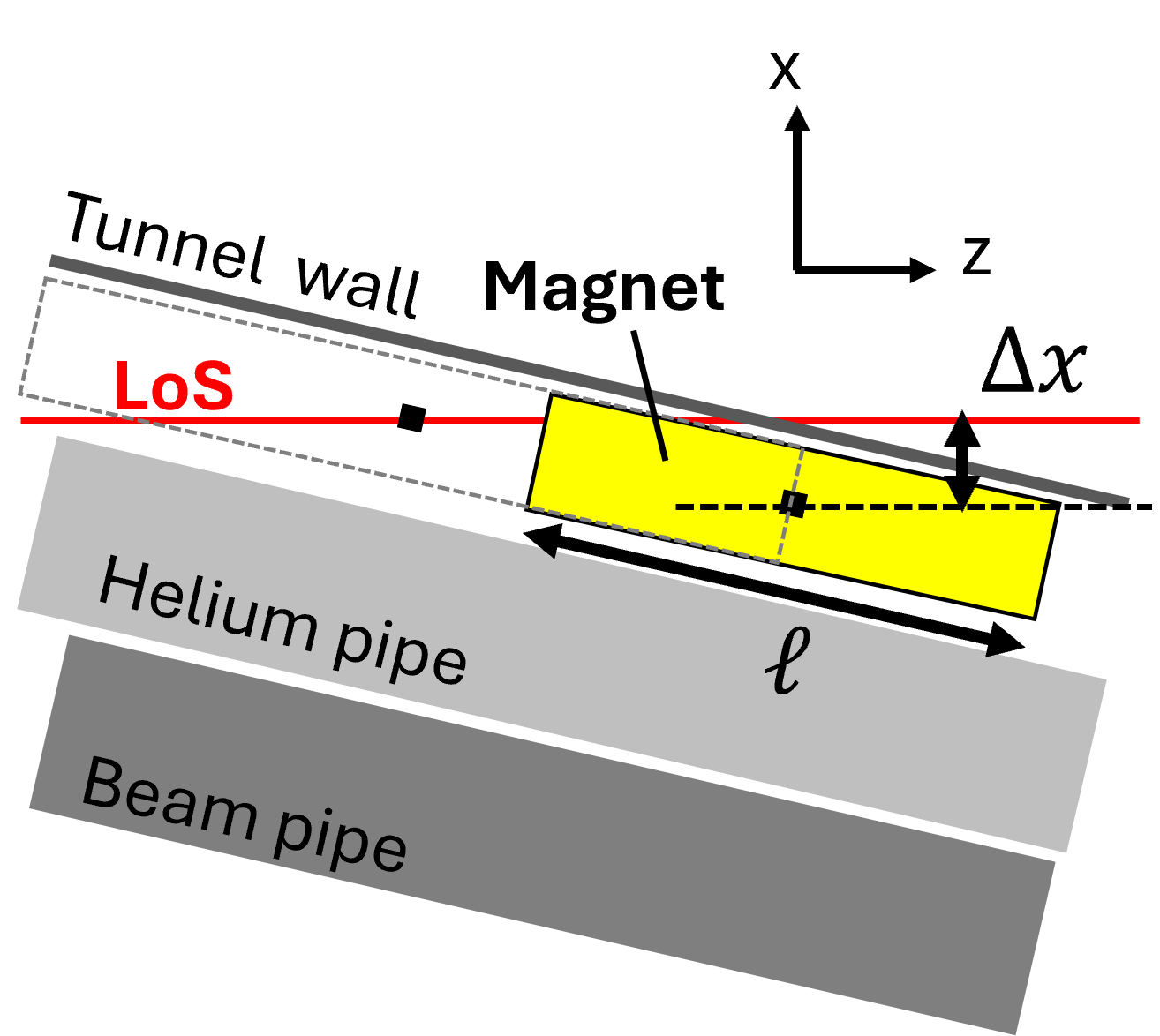}
        \includegraphics[width=0.35\linewidth]{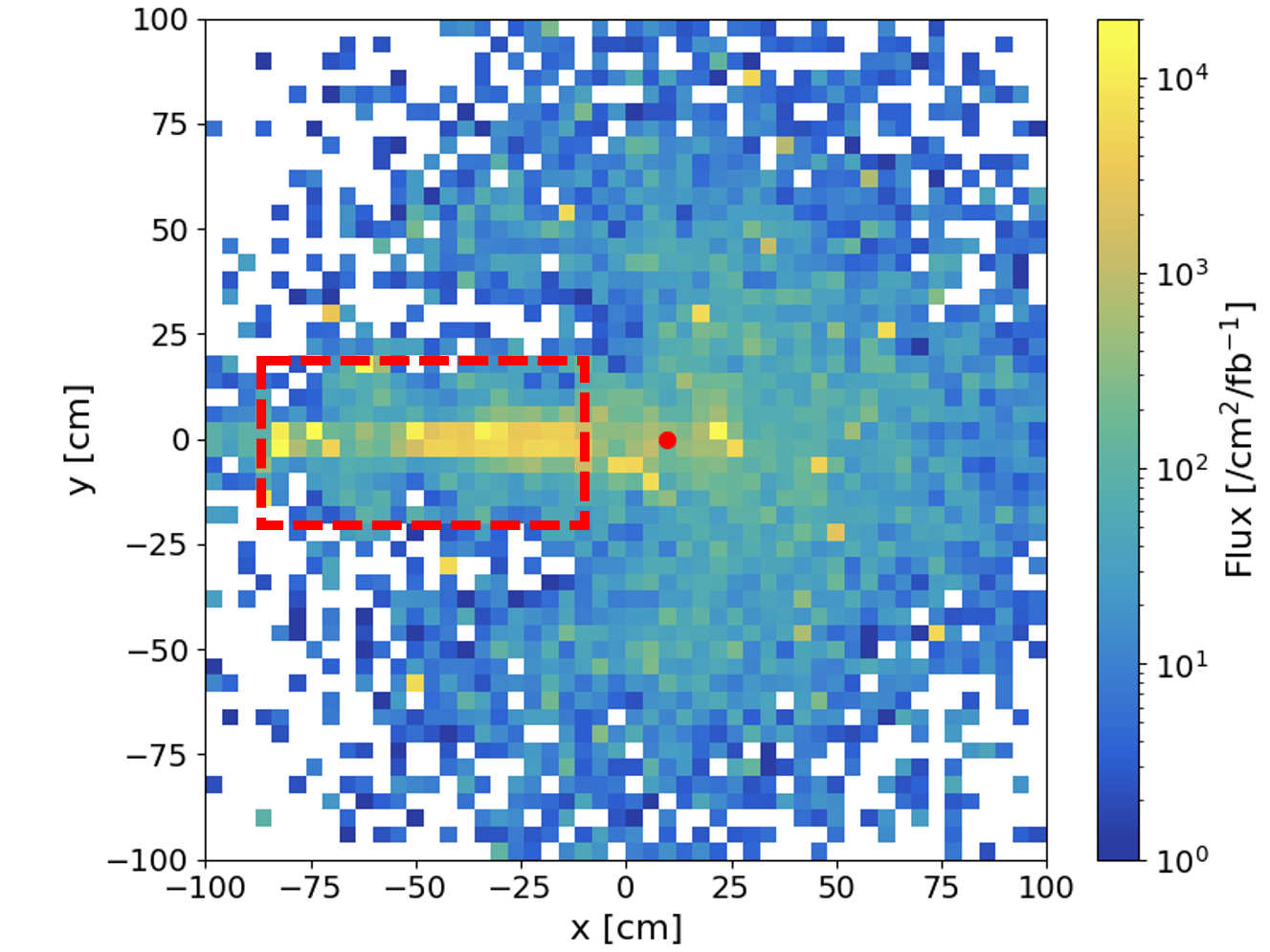}
        \includegraphics[width=0.35\linewidth]{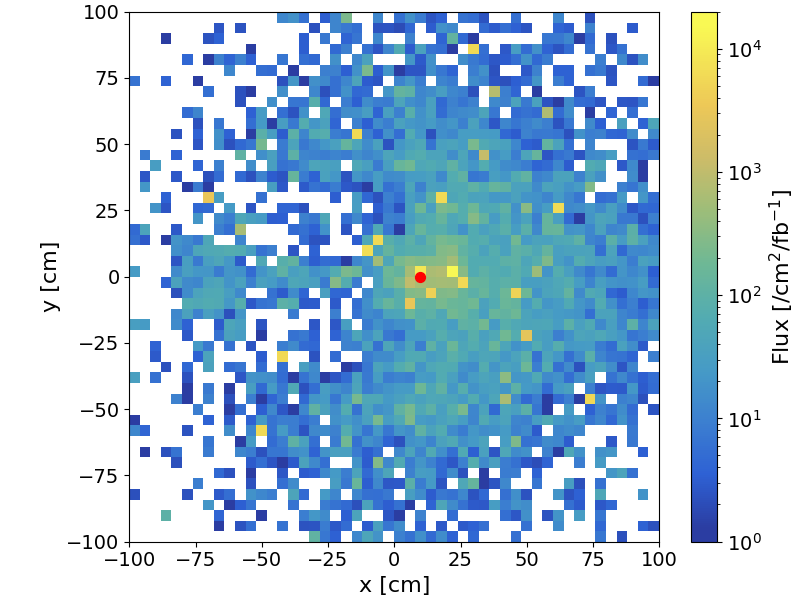}
\caption{
Left: schematic illustration of the definitions of the magnet length $\ell$ and transverse offset $\Delta x$ with respect to the LoS. Negative values of $\Delta x$ correspond to magnet positions closer to the beam pipe. Center: transverse muon distribution at the 370~m location for muons reaching the detector in the absence of a sweeper magnet. The red box indicates the coverage of the $40~\mathrm{cm} \times 40~\mathrm{cm}$ magnet configuration with $(\ell, \Delta x) = (27~\mathrm{m}, -50~\mathrm{cm})$, and the red marker indicates the LoS position. Right: the same distribution as in the center panel, but with the LHC-B sweeper magnet configuration with $(\ell, \Delta x) = (27~\mathrm{m}, -50~\mathrm{cm})$ applied.
}
\label{fig:dxell}
\end{figure}

The magnets are installed parallel to the tunnel axis, as illustrated in Fig.~\ref{fig:dxell}. Three magnet lengths, $\ell = 18$, 27, and 36~m, are considered together with several transverse offsets relative to the LoS. Figure~\ref{fig:dxell} also shows the transverse muon distribution at the 370~m location for muons reaching the detector, both in the absence of a sweeper magnet and for the LHC-B configuration with $(\ell, \Delta x) = (27~\mathrm{m}, -50~\mathrm{cm})$ as an example.

A clear reduction in the muon population passing through the magnet region is observed when the sweeper magnet is applied. However, the suppression performance is intrinsically limited by the broad transverse spread of the muon distribution, since a finite-size magnet cannot cover the entire phase-space region occupied by the background muons.

\begin{figure*}[t]
    \centering

    \includegraphics[width=\linewidth]{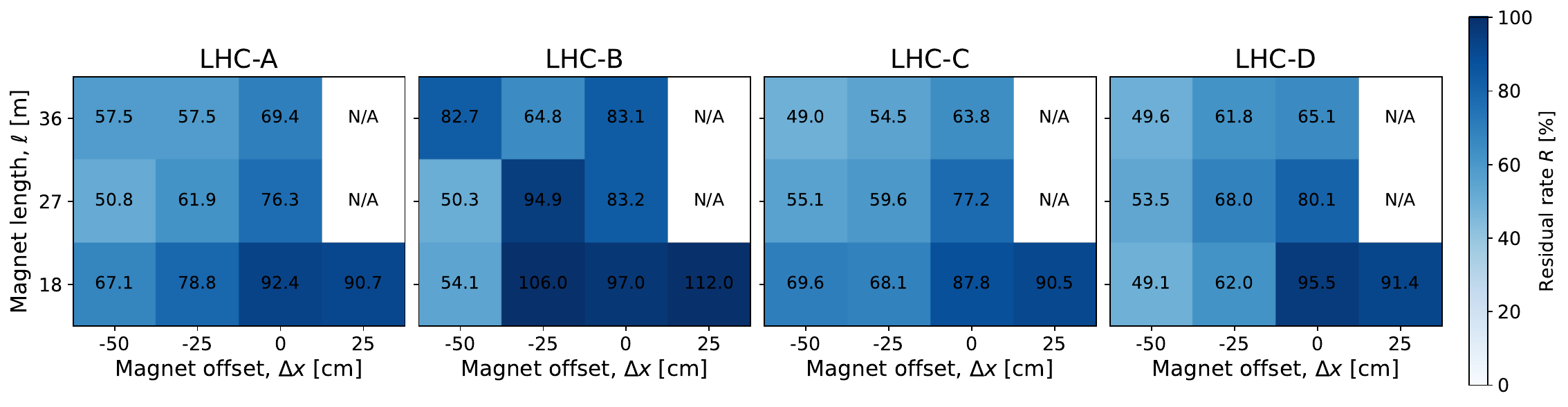}

    \caption{
    Residual muon fraction $R$ for different combinations of magnet length $\ell$ and transverse offset $\Delta x$ for the four LHC tunnel magnet configurations. The entry ``N/A'' indicates configurations that were not evaluated. The negative values of $\Delta x$ correspond to magnet positions closer to the beam pipe.
    }
    \label{fig:rate_lhc}
\end{figure*}

Figure~\ref{fig:rate_lhc} shows the residual ratio $R$ for the different magnet configurations and placement conditions. In general, configurations shifted toward negative $x$ achieve stronger suppression than configurations centered on the LoS. 
This behavior reflects the asymmetric muon distribution produced by the LHC lattice and downstream beamline geometry, which favors magnet placements shifted toward the beam-pipe side.
Longer magnets also tend to provide improved suppression.

Representative optimized configurations for each magnet type are summarized in Table~\ref{tab:flux_mag_lhc}. The resulting muon flux is reduced to approximately $2\times10^3~\mathrm{cm^{-2}}/\mathrm{fb}^{-1}$, comparable to the target flux required for annual replacement of the FASER$\nu$2 emulsion detector.

For the LHC-A and LHC-B configurations, additional studies were performed using smaller magnets with half the transverse dimensions. Although these compact configurations still provide a measurable suppression effect, their performance is significantly worse than that of the larger magnet configurations, indicating that sufficient transverse coverage is important for effective muon mitigation.

\begin{table}[t]
    \centering
    \caption{
    Muon fluxes and residual fractions at the FASER$\nu$2 detector location for different magnet configurations. The residual fraction $R$ is defined relative to the no-magnet configuration.
    }
    \label{tab:flux_mag_lhc}

    \begin{tabular}{cccccc}
        \hline
        Configuration & Magnet size [cm$\times$cm] & $\ell$ [m] & $\Delta x$ [cm] & Muon flux $\Phi$ [$10^3~\mathrm{cm^{-2}}/\mathrm{fb}^{-1}$] & $R$ \\
        \hline
        No magnet & -- & -- & -- & $3.76 \pm 0.32$ & $1.00$ \\
        \hline
        LHC-A & $40 \times 40$ & 27 & -50 & $1.91 \pm 0.23$ & $0.51 \pm 0.08$ \\
        LHC-B & $40 \times 40$ & 27 & -50 & $1.89 \pm 0.24$ & $0.50 \pm 0.08$ \\
        LHC-C & $40 \times 80$ & 36 & -50 & $1.84 \pm 0.23$ & $0.49 \pm 0.08$ \\
        LHC-D & $40 \times 80$ & 18 & -50 & $1.84 \pm 0.23$ & $0.49 \pm 0.08$ \\
        \hline
        LHC-A & $20 \times 20$ & 36 & -25 & $2.53 \pm 0.28$ & $0.67 \pm 0.09$ \\
        LHC-B & $20 \times 20$ & 36 & -25 & $2.30 \pm 0.27$ & $0.61 \pm 0.09$ \\
        \hline
    \end{tabular}
\end{table}

\begin{figure}[t]
    \centering

        \centering
        \includegraphics[width=0.4\linewidth]{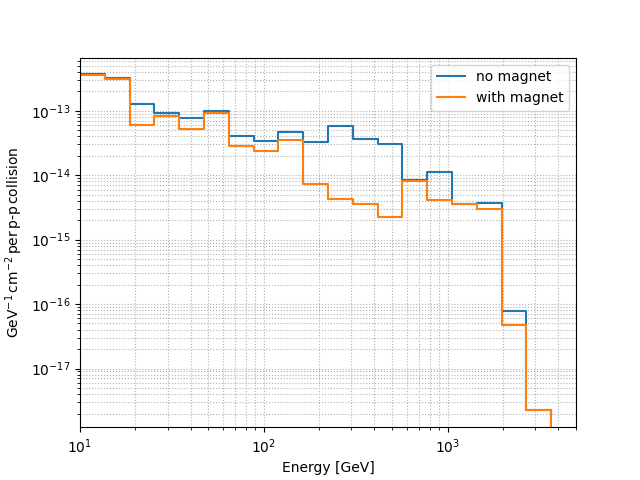}
        \includegraphics[width=0.4\linewidth]{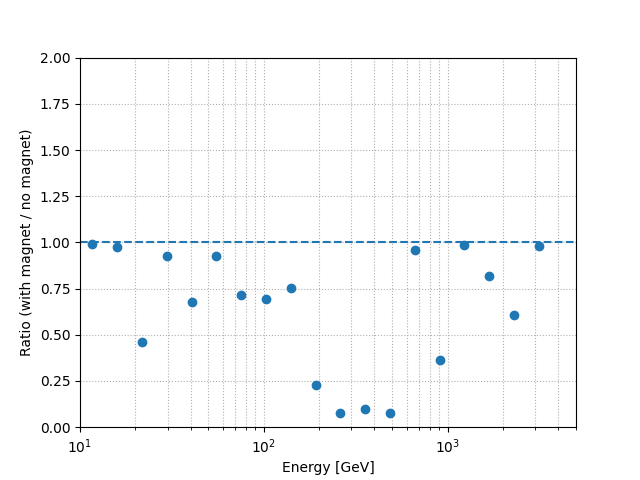}

    \caption{
    Comparison of the muon energy distributions for the representative LHC-A magnet configuration with and without the sweeper magnet. The left panel shows the energy distributions for the no-magnet case (blue) and the magnet configuration (orange). The right panel shows the ratio of the distributions with and without the sweeper magnet as a function of muon energy. The statistical fluctuations visible in the distributions originate from the mixture of events with significantly different event weights.
    }
    \label{fig:lhc-a_energy}
\end{figure}

Figure~\ref{fig:lhc-a_energy} compares the muon energy spectra before and after the installation of the LHC-A configuration. The suppression is most significant in the energy range between 100~GeV and 1~TeV.

\subsection{Combined configuration: LHC tunnel and TI18 tunnel}

To further reduce the background, we investigate configurations in which additional magnets are installed in the TI18 tunnel and at the entrance of the FPF.

We first consider the case where an additional magnet with a length of 1~m is installed in the TI18 tunnel. The geometrical constraints on the magnet placement in TI18 requires $x \leq x_{\mathrm{LoS,TI18}}$ and $y \geq 0$. Therefore, one corner of the magnet cross section is fixed at $(x, y) = (x_{\mathrm{LoS,TI18}}, 0)$, and different transverse sizes are evaluated.

Table~\ref{tab:flux_mag_ti18} summarizes the muon flux for representative configurations, where the LHC-A magnet ($( \ell, \Delta x ) = (27~\mathrm{m}, -50~\mathrm{cm})$) is combined with TI18-A magnets of size $20 \times 20$~cm$^2$ and $40 \times 40$~cm$^2$. The resulting residual fraction remains comparable to the case with only the LHC-A magnet ($0.51 \pm 0.08$), showing no significant improvement. This is mainly due to the limited magnet length of 1~m imposed by spatial constraints.

\begin{table}[t]
    \centering
    \caption{
    Muon fluxes and residual fractions for configurations with an additional TI18 magnet. The residual fraction $R$ is defined relative to the no-magnet case.
    }
    \label{tab:flux_mag_ti18}
    \begin{tabular}{ccccc}
        \hline
        Configuration & Width [cm] & Height [cm] & $\Phi$ [$10^3~\mathrm{cm^{-2}}/\mathrm{fb}^{-1}$] & $R$ \\
        \hline
        LHC-A only & -- & -- & $1.91 \pm 0.23$ & $0.51 \pm 0.08$ \\
        LHC-A + TI18-A & 20 & 20 & $1.82 \pm 0.22$ & $0.48 \pm 0.08$ \\
        LHC-A + TI18-A & 40 & 40 & $1.86 \pm 0.24$ & $0.49 \pm 0.08$ \\
        LHC-A + TI18-B & 80 & 50 & $1.63 \pm 0.20$ & $0.43 \pm 0.07$ \\
        \hline
    \end{tabular}
\end{table}

\begin{figure}[t]
    \centering
    \includegraphics[width=0.4\linewidth]{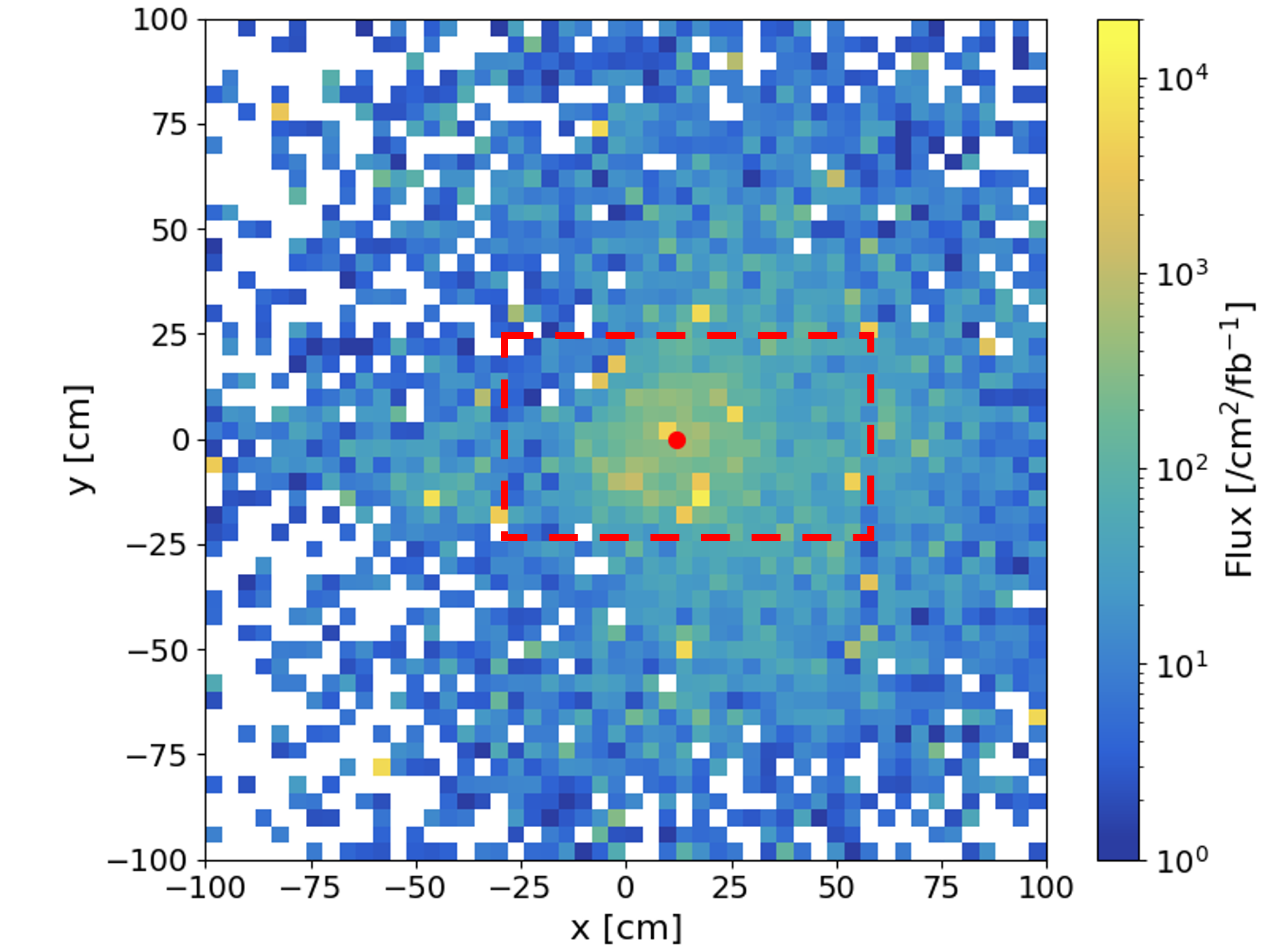}
    \includegraphics[width=0.4\linewidth]{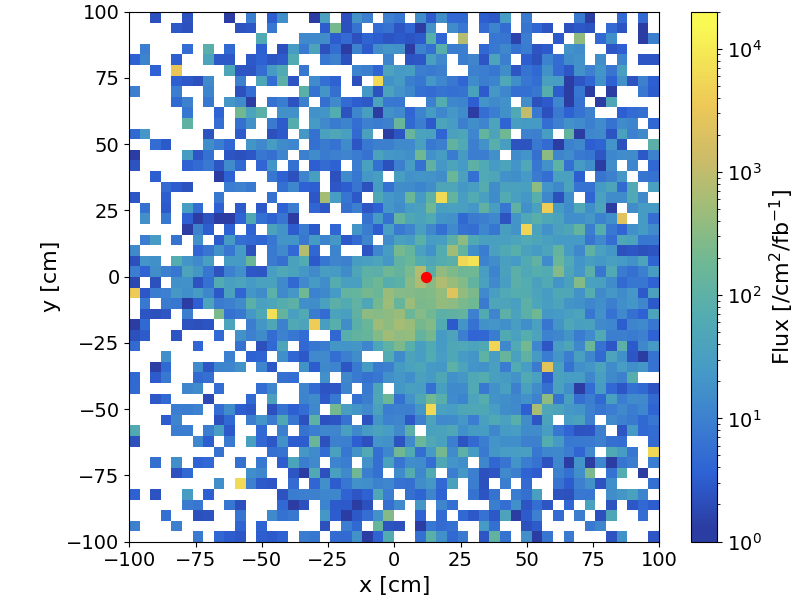}
    \caption{
    Transverse muon distributions at the TI18 location (480~m) for muons reaching the detector. Left: LHC-A only. Right: LHC-A with an additional TI18-B magnet placed on the LoS.
    }
    \label{fig:ti18_removesnd}
\end{figure}

For comparison, a hypothetical configuration without spatial constraints is also considered, in which a larger magnet (TI18-B) with a length of 3~m and size $80 \times 50$~cm$^2$ is placed on the LoS. This configuration corresponds to a scenario where the tunnel floor is excavated. In this case, the residual fraction is reduced to $0.43 \pm 0.07$, indicating a moderate additional suppression.
The transverse muon distributions at the TI18 location shown in Fig.~\ref{fig:ti18_removesnd} indicate that the muon flux is reduced within the magnet region.

\subsection{Combined configuration: LHC tunnel and FPF entrance}

We next consider the case where an additional magnet with a length of 1~m is installed at the entrance of the FPF. The transverse dimensions of the magnet are chosen to match the detector size of $25~\mathrm{cm} \times 64~\mathrm{cm}$, which also approximately corresponds to the transverse extent of the muon distribution at this location. In all configurations, the magnet is centered on the LoS.

For the representative configuration (FPF-A), the resulting muon flux is $(1.69 \pm 0.21)\times10^{3}~\mathrm{cm^{-2}}/\mathrm{fb}^{-1}$, corresponding to a residual fraction of $R = 0.45 \pm 0.07$. Although the magnet is located only 14~m upstream of the detector, a noticeable suppression is achieved.

This suppression is primarily attributed to the reduction of low-energy muons, which are more effectively deflected even over a short distance. In contrast, higher-energy muons are less affected due to their larger rigidity, limiting the overall suppression performance of this configuration.

\subsection{Combined configuration: all three locations}

The results for the combined magnet configurations are summarized in Table~\ref{tab:summary_all}. The LHC-A configuration alone already achieves the target muon flux of $2\times10^3~\mathrm{cm^{-2}}/\mathrm{fb}^{-1}$ required for the operation of the FASER$\nu$2 emulsion detector. 

By installing the magnets at all three locations, the muon flux is further reduced to $1.54\times10^3~\mathrm{cm^{-2}}/\mathrm{fb}^{-1}$, corresponding to a remaining rate of $0.41$. This demonstrates that a multi-stage magnet system provides additional suppression, although the improvement is moderate compared to the primary upstream magnet.

\begin{table}[t]
    \centering
    \caption{
    Summary of muon flux and residual rate for different magnet configurations. The residual rate $R$ is defined with respect to the no-magnet case. Here, the LHC-A configuration corresponds to $(\ell, \Delta x) = (27~\mathrm{m}, -50~\mathrm{cm})$.
    }
    \label{tab:summary_all}

    \begin{tabular}{lcc}
        \hline
        Configuration & $\Phi$ [$10^3~\mathrm{cm^{-2}}/\mathrm{fb}^{-1}$] & $R$ \\
        \hline
        No magnet & $3.76 \pm 0.32$ & $1$ \\
        (1) LHC-A & $1.91 \pm 0.23$ & $0.51 \pm 0.08$ \\
        (1) LHC-A + (2) on LoS & $1.63 \pm 0.20$ & $0.43 \pm 0.07$ \\
        (1) LHC-A + (3) & $1.69 \pm 0.22$ & $0.45 \pm 0.07$ \\
        (1) LHC-A + (2) on LoS + (3) & $\mathbf{1.54 \pm 0.20}$ & $\mathbf{0.41 \pm 0.07}$ \\
        \hline
    \end{tabular}
\end{table}

\begin{figure}
    \centering
    \begin{tabular}{ccc}
    \includegraphics[width=0.32\linewidth]{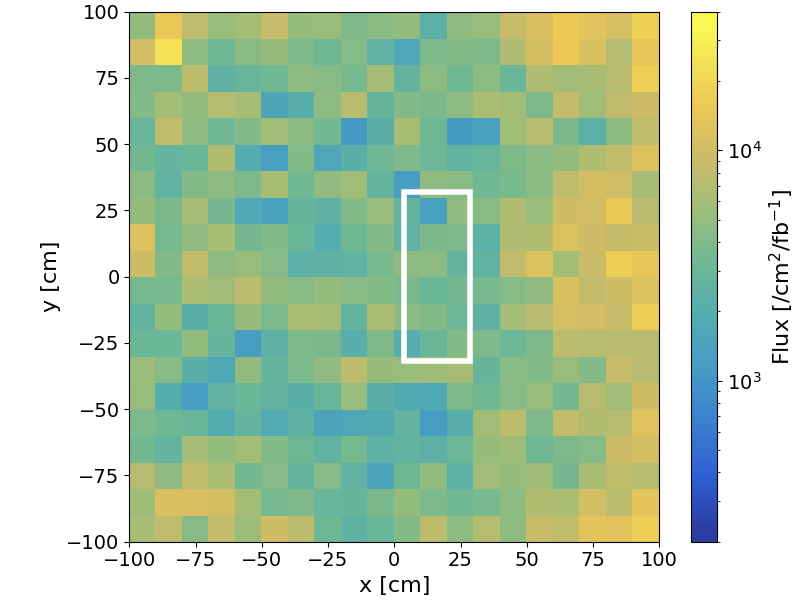} &
    \includegraphics[width=0.32\linewidth]{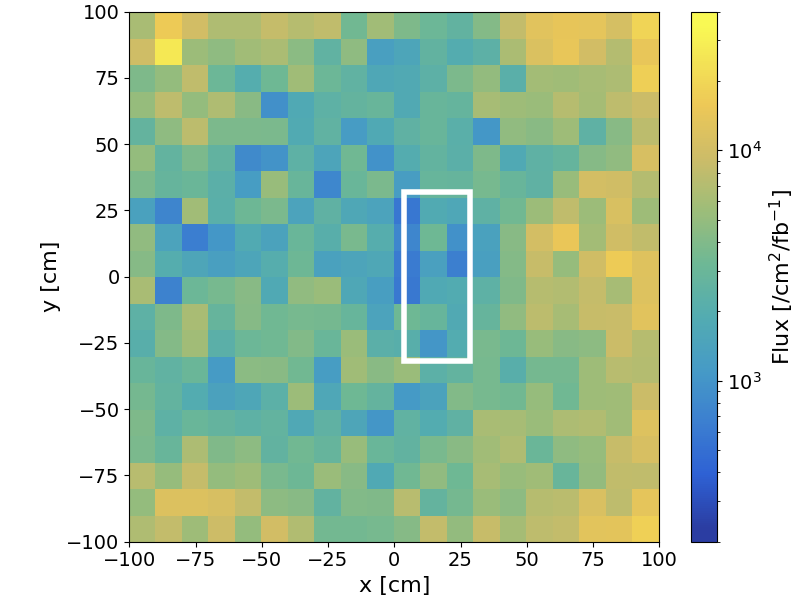} &
    \includegraphics[width=0.32\linewidth]{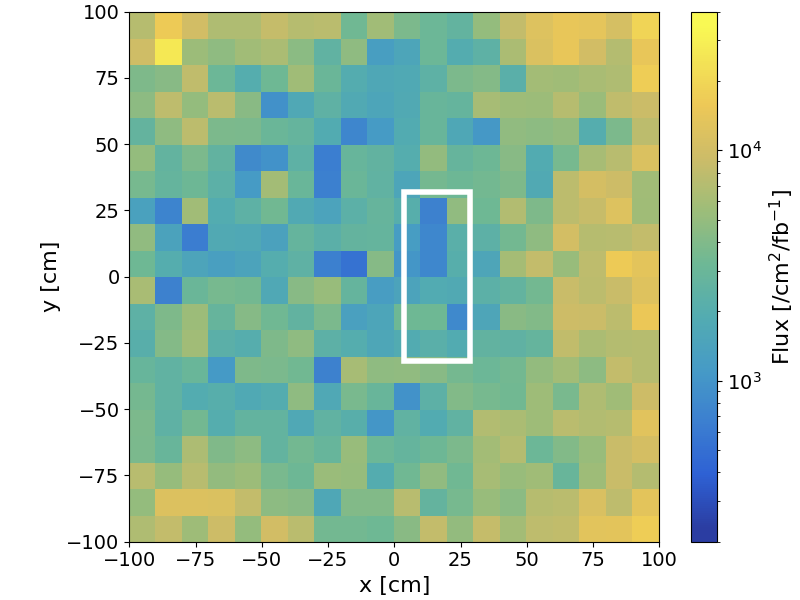} \\
    No magnet & (1) LHC-A & (1) LHC-A + (2) on LoS + (3)\\
    \end{tabular}
    \caption{Transverse muon profiles at the FASER$\nu$2 detector location for the no-magnet configuration (left), the LHC-A-only configuration (center), and the combined three-location magnet configuration (right). The white rectangle indicates the FASER$\nu$2 detector acceptance centered on the LoS.}
    \label{fig:muonprofile_FASERnu2}
\end{figure}

The resulting transverse muon profiles at the FASER$\nu$2 location are shown in Figure~\ref{fig:muonprofile_FASERnu2}. A significant reduction of the muon density around the LoS is observed after the application of the sweeper magnet system. While the upstream LHC-tunnel magnet provides the dominant suppression effect, the additional downstream magnets further reduce the residual muon population near the detector acceptance.

\section{Conclusion}

In this study, we investigated the suppression of background muons for the FASER$\nu$2 experiment using sweeper magnet systems installed upstream of the detector. A comprehensive simulation framework was developed by combining SIBYLL event generation, BDSIM beam transport, Geant4 particle tracking, and realistic magnetic field maps calculated with finite element methods.

A conceptual study based on a toy Monte Carlo simulation demonstrated that the interplay between magnetic deflection and multiple Coulomb scattering plays a key role in determining the muon suppression performance. In particular, the study revealed the existence of an energy-dependent optimal distance between the magnet and the detector, providing important guidance for the magnet design.

Several magnet configurations and installation scenarios were evaluated at three representative locations: the LHC tunnel, the TI18 tunnel, and the entrance of the FPF. Among these, a tunnel-parallel magnet installed in the LHC tunnel provides the most effective suppression, already achieving the target muon flux of $2\times10^3~\mathrm{cm^{-2}}/\mathrm{fb}^{-1}$ required for the operation of the FASER$\nu$2 detector. Additional magnets installed at downstream locations further improve the suppression, reducing the muon flux to $1.54\times10^3~\mathrm{cm^{-2}}/\mathrm{fb}^{-1}$, corresponding to a remaining rate of $41\%$.

The results demonstrate that a properly optimized multi-stage sweeper magnet system can significantly mitigate the forward muon background in the HL-LHC forward region. At the same time, the limited additional gain from downstream magnets highlights the importance of early deflection and sufficient propagation distance. 

The current study is limited by the available muon statistics, leading to non-negligible uncertainties, and higher-statistics simulations will be required to refine the magnet optimization and support a more detailed engineering design. In addition, the position of the LoS can shift due to the beam crossing angle at the interaction point, introducing additional requirements on the alignment tolerance and robustness of the sweeper magnet system. Future studies incorporating alternative flux predictions, such as those based on FLUKA simulations, will also be important to assess model dependencies and improve the robustness of the results.

Overall, this work provides a realistic and quantitatively validated basis for the design of muon mitigation systems for future forward experiments at the FPF.

\section*{Acknowledgments}
\label{sec:Acknowledgments}
%%%%%%%%%%%%%%%%%%%%%%%%%%%%%%%%%%%%%%
%\input{acknowledgments}

This work was supported in part by ERC Consolidator Grant No.~101002690, Heising-Simons Foundation Grant Nos.~2018-1135, 2019-1179, and 2020-1840, Simons Foundation Grant No.~623683, and JSPS KAKENHI Grant Nos.~22H01233, 20K23373, 23H00103, and 25KJ0719.

The authors would like to thank Pierre Thonet of the CERN magnet group for valuable advice on the initial sweeper magnet design. We also acknowledge the Physics Beyond Colliders (PBC) study group and the CERN FLUKA team for their work on forward muon-flux studies for the Forward Physics Facility. In addition, we thank Julien Prosic and Jean-Pierre Corso for integration studies related to the placement of the sweeper magnet in the LHC tunnel.

%%%%%%%%%%%%%%%%%%%%%%%%%%%%%%%%%%%%%%
\bibliographystyle{utphys}
\bibliography{references}
%%%%%%%%%%%%%%%%%%%%%%%%%%%%%%%%%%%%%%

\end{document}